\documentclass[letterpaper]{article}

\usepackage{url}
\usepackage{jcd}
\usepackage{balance}
\usepackage[authoryear, round]{natbib}
\usepackage{xspace}
\usepackage{float}
\usepackage[utf8]{inputenc} % allow utf-8 input
\usepackage{fontenc}    % use 8-bit T1 fonts
\usepackage{url}            % simple URL typesetting
\usepackage{booktabs}       % professional-quality tables
\usepackage[font=small]{caption}
\usepackage[para,online,flushleft]{threeparttable}
\usepackage{amsfonts}       % blackboard math symbols
\usepackage{nicefrac}       % compact symbols for 1/2, etc.
\usepackage{microtype}      % microtypography
\usepackage{multicol}
\usepackage{enumitem}
\usepackage{xcolor}
\usepackage{placeins}
\usepackage{amsmath}
\usepackage{titlesec}
\usepackage{titletoc}
\usepackage{hyperref}       % hyperlinks
\usepackage{bbm}
\usepackage{times}
\usepackage{subfig}
\usepackage{alltt}
\usepackage{uai2019}
\usepackage[margin=1in]{geometry}

\newcommand{\ITT}{{\tt ITT}}

\newcommand{\iter}{{l}}
\newcommand{\unit}{i}

\newcommand{\x}{\mathbf{x}}
\newcommand{\vv}{\boldsymbol{\theta}}
\newcommand{\z}{\mathbf{z}}
\newcommand{\one}{\mathbf{1}}

\newcommand{\ind}{\mathbf{I}}

\newcommand{\MG}{{\tt MG}}
\newcommand{\mg}{{\tt{MG}}}

\newcommand{\Median}{{\tt Median}}

\definecolor{LightCyan}{rgb}{0.88,1,1}
\definecolor{Gray}{gray}{0.9}

\newcommand{\proj}[1]{{\Pi}}
\newcommand{\sel}[1]{{\sigma}}

\newcommand{\cut}[1]{}
\newcommand{\cutfull}[1]{}

\newcommand{\commentresolved}[1]{}

\newcommand{\ie}{{\it i.e.}} %\xspace}
\usepackage{aliascnt}  		% ``hyperref’s \autoref command does not work well with theorems that share a counter:
 \newaliascnt{lemma}{theorem}				% 1 alias counter
 \newtheorem{lemma}[lemma]{Lemma}              	% Lemma environment.
 \aliascntresetthe{lemma}  					% 3 set
\newtheorem{assumption}{Assumption}[section]

\cut{

}

\newcommand{\ismatched}{{\tt is\_matched}}

\newcommand{\BasicExactMatch}{{\tt GroupedMR}}

\newcommand{\air}{{AME-IV}}
\newcommand{\airr}{{AME-IV}}
\newcommand{\hair}{{\tt HAIRs }}
\newcommand{\MQ}{{\tt MQ}}

\newcommand{\BF}{{\tt BF}}
\newcommand{\PE}{{\tt PE}}

\newcommand{\ww}{\textbf{w}}
\newcommand{\MGi}{{\tt MG}(\boldsymbol{\theta}^{i*}, \x_i)}

\newcommand{\FLAMEIV}{{FLAME-IV}}
\newcommand{\FLAMEIVV}{FLAME-IV}

\title{\vspace{0.8cm} Interpretable Almost-Matching-Exactly With Instrumental Variables \\\vspace{0.5cm}}
\author{M. Usaid Awan\thanks{~~Equal contribution.},~ Yameng Liu\footnotemark[1],~~Marco Morucci\footnotemark[1],~~ Sudeepa Roy$^{}$, Cynthia Rudin$^{}$, Alexander Volfovsky$^{}$\\ %\vspace{.5cm} \\
Duke University\\
Durham, NC, USA
}

\begin{document}

\raggedbottom
\maketitle

\begin{abstract}
Uncertainty in the estimation of the causal effect in observational studies is often due to unmeasured confounding, i.e., the presence of unobserved covariates linking treatments and outcomes. Instrumental Variables (IV) are commonly used to reduce the effects of unmeasured confounding. Existing methods for IV estimation either require strong parametric assumptions, use arbitrary distance metrics, or do not scale well to large datasets. We propose a matching framework for IV in the presence of observed categorical confounders that addresses these weaknesses. Our method first matches units exactly, and then consecutively drops variables to approximately match the remaining units on as many variables as possible. We show that our algorithm constructs better matches than other existing methods on simulated datasets, and we produce interesting results in an application to political canvassing.
\end{abstract}
\section{INTRODUCTION}
\label{sec:intro}
% \textcolor{red}{NOTE: We need to add some discussion of how our method can be used to strengthen instruments to the paper.}
% \blfootnote{$^\dagger$Equal contribution. All authors listed alphabetically.}
The gold standard for inferring the causal effect of a treatment (such as smoking, a tax policy, or a fertilizer) on an outcome (such as blood pressure, stock prices, or crop yield) is the randomized experiment: the analyst manually assigns the treatment to each of her units uniformly at random. Unfortunately, this manipulation is impossible or unethical for some treatments of practical interest, leading to the need for inferring causal relations from observational studies.
% In these settings researchers often resort to collecting data after treatments are pre-assigned non-uniformly to units. These types of studies are known as observational and are the focus of this paper. 
In many observational studies, it is common for \emph{instrumental variables (IV)} to be available. These variables are (a) allocated randomly across units, (b) correlated with the treatment, and (c) affect the dependent variable only through their effect on the treatment. The fact that instrumental variables allow for consistent estimation of causal effect with non-randomized treatments is a hallmark of the causal inference literature, and has led to the use of IV methods across many different applied settings \citep[e.g.,][]{joskow1987contract, gerber2000, acemoglu2001, autor2013}.

The most popular existing method that uses instrumental variables to conduct causal inference is Two-Stage Least Squares Regression (2SLS) \citep{angrist1991does,card1993using,wooldridge2010econometric}. The 2SLS methodology makes strong parametric assumptions about the underlying outcome model (linearity), which do not generalize well to complex problems. Non-parametric approaches to IV-based causal estimates generalize 2SLS to more complex models \citep{newey2003instrumental, frolich2007}, but lack interpretability; it is difficult to troubleshoot or trust black box models.
Matching methods that allow for nonparametric inference on average treatment effects without requiring functional estimation have recently been introduced for the IV problem in \cite{kang2016full}:  the full-matching algorithm presented in their work relaxes some of the strong assumptions of 2SLS, however, it does not scale well to massive datasets, and imposes a fixed metric on covariates. It also does not take into account that covariates have different levels of importance for matching. 
%\textcolor{red}{THIS NEEDS TO BE CHECKED} 

The approach for instrumental variable analysis presented in this paper aims to handle the problems faced by existing methods: it is non-parametric, scalable, and preserves the interpretability of having high-quality matched groups.  We create an Almost-Matching Exactly framework  \citep{roy2017flame, dieng2018collapsing} for the purpose of instrumental variable analysis. Our methodology estimates the causal effects in a non-parametric way and hence performs better than 2SLS  or other parametric models. It improves over existing matching methods for instrumental variables when covariates are discrete, leveraging an adaptive distance metric. This adaptive distance metric is capable of systematically accounting for nuisance variables, discounting their importance for matching. The algorithm scales easily to large datasets (millions of observations) and can be implemented within most common database systems for optimal performance. 

In what follows, first we introduce the problem of instrumental variable estimation for observational inference, and describe the role of matching within it. Second, we outline the Almost-Matching Exactly with Instrumental Variables (\air) framework for creating matched groups. Third, we describe estimators with good statistical properties that can be used on the matched data. Finally, we present results from applying our methodology to both simulated and real-world data: we show that the method performs well in most settings and outperforms existing approaches in several scenarios. 
%Our main contribution is to extend the framework of alm ost-exact mat (AME) to instrumental variables (IV), where the new framework is called AME-IV.

\section{RELATED WORK}
%Instrumental variables(IVs) is a well known approach in literature to estimate the causal effect of exposure on the outcome when there is uncontrolled confounding.To measure the causal effect, an instrumental variable has to be valid. An valid instrument has to satisfy (A1)associate with exposure, (A2)only affects outcome through its effect on exposure, and (A3)is not associated with any unmeasured confounders after controlling all measured confounders\citep{angrist1996identification,baiocchi2014instrumental}. An instrumental variable should also has strong association with exposure, because a weak instrument is invariably sensitive to very small nonrandom assignment of the instruments\citep{small2008war}.\\
Widely used results on definition and identification of IVs are given in \cite{imbens1997bayesian, angrist1996identification}, and generalized in \cite{brito2002generalized, chen2016incorporating}. Methods for discovery of IVs 
% in observational data 
are developed in \cite  {silva2017}.

The most popular method for IV estimation in the presence of observed confounders is two-stage least squares (2SLS) \citep{card1993using}. 2SLS estimators are consistent and efficient under linear single-variable structural equation models with a constant treatment effect \citep{wooldridge2010econometric}. One drawback of 2SLS is its sensitivity to misspecification of the model. Matching, on the other hand, allows for correct inference without the need to specify an outcome model. 
%Recent work on matching for IV estimation includes propensity score matching \citep{ichimura2001propensity}, \textcolor{red}{how do these relate to our work? Why is our work novel?}pair matching \citep{baiocchi2010building}, matching with fixed number of units with level 0 of the IV \citep{kang2013causal} and full matching \citep{kang2016full}. These methods are constrained to depend on continuous distance metrics between covariates and can lead to faulty inferences in categorical data settings. 

Recent work on matching for IV estimation includes matching methods that match directly on covariates, rather than on summary statistics like propensity score \citep{ichimura2001propensity}. These matching methods can be very powerful nonparametric estimators; full matching \citep{kang2013causal} is one such approach, but has a limitation in that its distance metric between covariates is fixed, whereas ours is learned. \cite{roy2017flame} provides an in-depth discussion of other matching methods including near-far and full-matching, in the context of AME.

Other IV methods in the presence of measured covariates include Bayesian methods \citep{imbens1997bayesian}, semiparametric methods  \citep{abadie2003semiparametric,tan2006regression,ogburn2015doubly}, nonparametric methods \citep{frolich2007} and deep learning methods \citep{hartford2017deep},  but these methods do not enjoy the benefits of interpretability that matching provides. 
%In contrast, in this paper we proposed a novel almost-matching method that uses weighted hamming distance metrics and generates inference results with high quality with tradeoff to computational speed.
\section{METHODOLOGY}\label{methodology}

We consider the problem of instrumental variable estimation for a set of $n$ units indexed by $i=1,\dots, n$. Each unit is randomly assigned to a binary instrument level. Units respond to being assigned different levels of this instrument by either taking up the treatment or not: we denote with $t_{i}(1),t_{i}(0) \in \{0, 1\}$ the treatment level taken up by each unit after being exposed to value $z \in \{0,1\}$ of the instrument. Subsequently, units respond to a treatment/instrument regime by exhibiting different values of the outcome variable of interest, which we denote by $y_{i}(t_{i}(1), 1),y_{i}(t_{i}(0), 0) \in \mathbb{R}$. Note that this response depends both on the value of the instrument assigned (2nd argument) and on the treatment value that units  take up in response to that instrument value (1st argument). All quantities introduced so far are fixed for a given unit $i$ but not always observed. In practice, we have a random variable $Z_{i} \in \{0, 1\}$ for each unit denoting the level of instrument that it was assigned, and observed realizations of $Z_i$ are denoted with $z_i$. Whether a unit receives treatment is now a random variable ($T_{i}$), and the outcome is random ($Y_{i}$), and they take the form: 
\begin{align*}
Y_{i} &= y_{i}(t_{i}(1), 1)Z_{i} + y_{i}(t_{i}(0), 0)(1-Z_{i})\\
T_{i} &= t_{i}(1)Z_{i} + t_{i}(0)(1-Z_{i}).
\end{align*}
Note that the only randomness in the observed variables comes from the instrument, all other quantities are fixed. We use $y_i$ and $t_i$ to denote observed realizations of $Y_i$ and $T_i$ respectively. We also observe a fixed vector of $p$ covariates for each unit, $\x_i \in \mathcal{X}$, where $\mathcal{X}$ is a space with $p$ dimensions. In this paper we are interested in the case in which $\mathcal{X} = \{0, 1\}^p$, corresponding to categorical variables, where exact matching is well-defined. 

\cut{For convenience, "treatment" and "control" in the left paper represents units with $Z_i$ = 1 or 0. Consider data $D = [y,x,t,z]$ where $Y\in \mathbb{R}^n$, $X\in \{0,1\}^{n\times p}$, $T\in\{0,1\}^n$, $Z \in \{0,1\}^n$ respectively denote the outcome vector, the covariates for all units, the treatment indicator (1 for treated, 0 for control) and instrumental variable. The $j$-th covariate $X$ for unit $i$ is denoted $x_{ij} \in \{0,1\}$. Notation $\mathbf{x}_i \in \{0,1\}^p$ indicates covariates for the $i$th unit, $T_i \in \{0,1\}$ is an indicator for whether or not unit $i$ is treated and $Z_i \in \{0,1\}$ indicates the value associated with the instrument variable for unit $i$. \\}

Throughout we make the SUTVA assumption, that is (i) outcome and treatment assignment for each individual are unrelated to the instrument exposure of other individuals, and (ii) the outcome for each individual is unrelated to the treatment assignment of other individuals \citep{angrist1996identification}. However, ignorability of treatment assignment is not required. We make use of the instrumental variable to estimate the causal effect of treatment on outcome. In order for a variable to be a valid instrument it must satisfy the following standard assumptions \citep[see, e.g.,][]{imbens1994identification, angrist1996identification, imbens2015}:

\noindent \textbf{(A1) Relevance}: $\frac{1}{n}\sum_{i=1}^n t_{i}(1) - t_{i}(0) \neq 0$, that is, the variable does indeed have a non-zero causal effect on treatment assignment, on average. 

\noindent \textbf{(A2) Exclusion}: If $z \neq z' \mbox{ and } t_{i}(z) = t_{i}(z') $ then $y_{i}(t_{i}(z), z) = y_{i}(t_{i}(z'), z') $ for each unit $i$. This assumption states that unit $i$'s potential outcomes are only affected by the treatment it is exposed to, and not by the value of the instrument. \cut{If we were to manually assign a treatment to the units, that the instrument would have no effect on them. }Therefore, $y_{i}(t_{i}(z), z)$ can be denoted by:  $y_{i}(t_{i}(z))$. 

\noindent \textbf{(A3) Ignorability}: $Pr(Z_{i} = 1|\x_i) = e(\x_i)$ for all units $i$, and some non-random function $e: \mathcal{X} \mapsto (0,1)$. This assumption states that the instrument is assigned to all units that have covariate value $\x_i$ with the same probability. It implies that if two units $i$ and $k$ have $\x_i = \x_k$, then $\Pr(Z_i = 1 | \x_i) = \Pr(Z_k = 1|\x_k)$. 

\noindent \textbf{(A4) Strong Monotonicity}: $ t_{i}(1) \geq t_{i}(0)$ for each unit $i$. \cut{This assumption states that the same unit cannot take up the treatment if it does not receive the instrument and not take up the treatment if it receives the instrument.} This assumption states that the instrument is seen as an encouragement to take up the treatment, this encouragement will only make it more likely that units take up the treatment and never less likely. 

\begin{figure}[t]
    \centering
    \includegraphics[width=.35\textwidth]{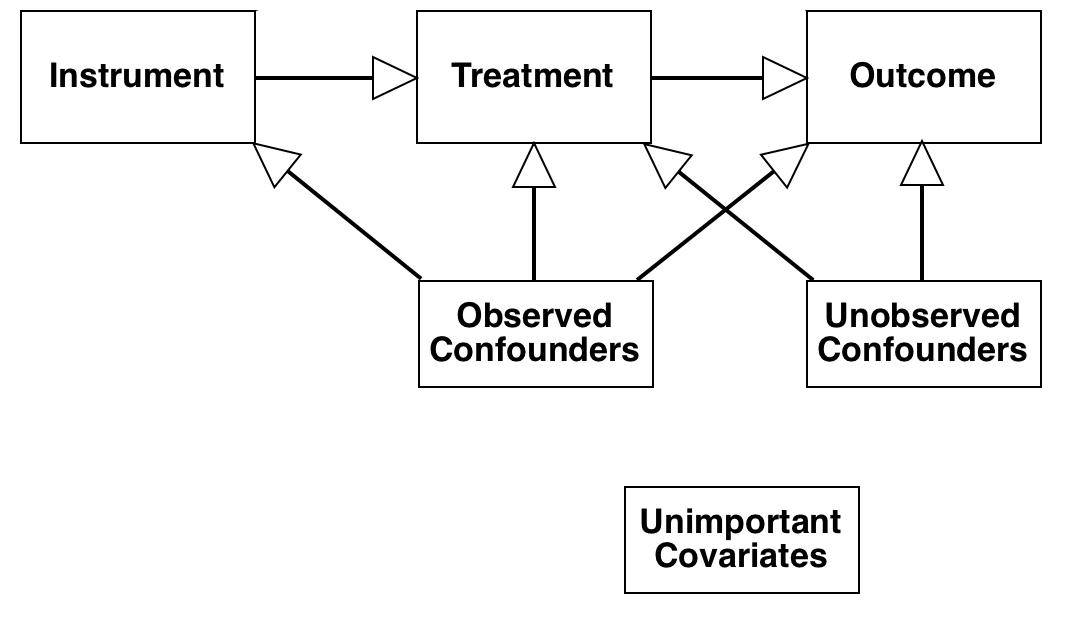}
    \caption{Causal DAG for instrumental variables. Arrows represent causal relationships between variables. The lack of a direct arrow from instrument to outcome represents Assumption A2 and the lack of a direct arrow from unobserved confounders to the instrument represents A3. \label{fig:unit_effect}}
\end{figure}

Figure \ref{fig:unit_effect} gives a graphical summary of the identification assumptions. An instrumental variable satisfying (A1, A2, A3 and A4) allows us to estimate the treatment effect for a subgroup that responds positively to exposure to the instrument \citep{imbens1994identification}. We note that these are not the only criteria for the use of instrumental variables, for example \cite{brito2002generalized} introduces a graphical criterion for identification with instrumental variables. \cut{For example, if $z$ reflects assignment to treatment and $t$ reflect whether or not treatment was undertaken, we estimate the causal effect of $t$ on $y$ for the individuals for whom $z$ was instrumental in deriving the decision to undertake the treatment $t$.} These are units that would have undertaken the treatment only after administration of the instrument and never without \citep{angrist1996identification}. Note that we cannot identify these units in our sample, given what we observe, but we can estimate the treatment effect on them \citep{imbens2015}. This treatment effect is known as Local Average Treatment Effect (LATE) and takes the following form \citep{imbens1994identification, angrist1996identification}:
\begin{align}
\lambda &= \frac{1}{n_c}\sum_{\substack{i: t_{i}(1) > t_{i}(0)}} y_{i}(1) - y_{i}(0)\nonumber\\ 
&=\frac{\sum_{\x \in \mathcal{X}} \omega_\x \ITT_{y, \x}}{\sum_{\x \in \mathcal{X}} \omega_\x \ITT_{t, \x}}\label{eq:late},
\end{align}
where $n_{c}$ is the total number of units such that $t_{i}(1) > t_{i}(0)$, $\omega_\x = n_\x/n$ is the weight associated with each value of $\x$, $n_\x$ is the number of units where $\x_i = \x$, and:
\begin{align*}
\ITT_{y, \x} &= \frac{1}{n_\x}\sum_{i: \x_i = \x} y_{i}(t_{i}(1)) -y_{i}(t_{i}(0))\\
\ITT_{t, \x} &= \frac{1}{n_\x}\sum_{i: \x_i = \x} t_{i}(1) - t_{i}(0).
\end{align*}
The quantities above are also known as the Intent-To-Treat effects: they represent the causal effects of the instrument on the outcome and the treatment, respectively. Intuitively, these effects can be estimated in an unbiased and consistent way due to ignorability of  instrument assignment (A3) conditional on units having the same value of $\x$. 

Approximate matching comes into this framework because in practice we almost never have enough treated and control units with the same exact values of $\x$ in our observed data to accurately estimate the quantities above. With approximate matching, we want to construct matched groups from observed $\x$ such that A3 holds approximately within each group. This means that a good approximate matching algorithm is one that produces groups where, if $i$ and $j$ are grouped together, then $\x_i \approx \x_j$. In the next section, we propose the Almost-Matching Exactly with Instrumental Variables (\air) framework to build good approximately matched groups from binary covariates. 
 
\subsection{ALMOST-MATCHING EXACTLY WITH INSTRUMENTAL VARIABLES (\airr\ PROBLEM)} 
The \air\ framework has the goal of matching each instrumented (i.e., $z_i = 1$) unit to at least one non-instrumented unit (i.e., $z_k = 0$) as exactly as possible. (The entire set of calculations is symmetric when we match each non-instrumented unit, thus w.l.o.g. we consider only instrumented units.) When units are matched on all covariates, this is an exact match. When units can be matched on the most important covariates (but not necessarily all covariates), this is an almost-exact match. The importance of covariate $j$ for matching is represented by a fixed nonnegative weight $w_j$. Thus, we consider the following problem for each instrumented unit $\unit$, which is to maximize the weighted sum of covariates on which we can create a valid matched group for $\unit$:
\begin{eqnarray*}
\lefteqn{\vv^{\unit*} \in \argmax\limits_{\vv\in\{0,1\}^p} \vv^T\mathbf{w}, \textit{ such that }} \\
&&\exists \;k \;\textit{ with }z_{k}=0 \textit{ and } \x_{k}\circ \vv = \x_{\unit}\circ\vv,  
\end{eqnarray*}
where $\circ$ denotes the Hadamard product, $\vv$ is a binary vector to represent whether or not each covariate is used for matching, and $\mathbf{w}$ is a nonnegative vector with a reward value associated with matching on each covariate. The constraint in our optimization problem definition guarantees that the main matched group of each instrumented unit $\unit$ contains at least one non-instrumented unit. The solution to this optimization problem is a binary indicator of the optimal set of covariates that unit $i$ can be matched on.
%For each instrumented unit $\unit$, we want to find such an optimal binary indicator $\vv^{\unit*}$.  
Note that, if all entries of $\vv^{i*}$ happen to be one, then the units in unit $i$'s main matched group will be exact matches for $i$. 

We define $\unit$'s \textbf{main matched group} in terms of $\vv^{\unit*}$ as:
$$\MGi=\{ k:\vv^{\unit*}\circ\x_k=\vv^{\unit*}\circ \x_{\unit}\}.$$ 
We now theoretically connect Assumption A3 with solving the \air\ problem, and show how approximate matches can lead to the assumption being approximately satisfied within each matched group. This makes IV estimation possible even when it is not possible to exactly match each unit. To do so, we introduce the notation $\one_{[\x_i \neq \x_k]}$ to denote a vector of length $p$ where the $j^{th}$ entry is one if $x_{ij} = x_{kj}$ and zero otherwise. 
\begin{lemma}\label{Theorem:Weighted Hamming Distance}
For any unit $\unit$ where $z_\unit$ = 1, with $\vv^{\unit*}$ as defined in the \air\ problem, then for any unit $k$ with $z_k \neq z_\unit$, if $\x_{k}\circ \vv^{\unit*} = \x_{\unit}\circ\vv^{\unit*}$, i.e., $k \in \MGi$, we have:
\begin{equation}
k \in \argmin\limits_{\substack{l = 1, \dots, n\\z_l \neq z_\unit}}\ww^T\one_{[\x_i \neq \x_l]}.
\end{equation}
In particular, if $\vv^{i*}$ has all entries equal to one and $k \in \MGi$ then $\ww^T\one_{[\x_i \neq \x_k]} = 0$.
\end{lemma}
The detailed derivation of this lemma is in the supplement. This statement clarifies that by solving the \air\ problem, we minimize the weighted hamming distance between each unit $\unit$ and all other units with a different assignment of the instrument that belong to $\unit$'s main matched group. We now introduce a smoothness assumption under which we can formally link the matched groups created by \air\ with the necessary conditions for causal estimation using instrumental variables.

%We call a weight vector $w$ a \textbf{smoothed weight} if for any two $i, k \in 1, \dots, n$, and $\delta>0$, we have: $\mathbf{w}^T(\x_i \xor \x_k) \leq \delta\implies |p(Z_i=1|\x_i)-p(Z_k=1|\x_k)| \leq \epsilon(\delta)$, where $\epsilon(\delta)$ is an increasing function of $\delta$.

\textbf{(A5) Smoothness:} For any two $\x_i, \x_k \in \{0,1\}^p$, and $\delta > 0 $, we have: $\mathbf{w}^T\one_{[\x_i\neq \x_k]} \leq \delta \implies |p(Z_i=1|\x_i)-p(Z_k=1|\x_k)|\leq\epsilon(\delta)$, where $\epsilon(\delta)$ is an increasing function of $\delta$ such that $\epsilon(0) = 0$.

Note that this is a variant of a standard assumption made in most matching frameworks \citep[see, e.g.,][]{rosenbaum2010design}.  The following proposition follows immediately from Lemma \ref{Theorem:Weighted Hamming Distance} applied to A5. 
\begin{proposition}\label{Lemma:Smooth}
If $k\in \MGi$ with $z_i \neq z_k$, and A5 holds, then $$|P(Z_i=1|\x_i) - P(Z_k=1|\x_k)| \hspace*{-2pt}\leq\hspace*{-2pt} \epsilon\hspace*{-2pt}\left(\hspace*{-2pt}\min\limits_{\substack{l=1\dots n\\ z_i \neq z_l}}\ww^T\one_{[\x_i \neq \x_l]}\hspace*{-2pt}\right)\hspace*{-2pt}.$$
In particular, if $\vv^{i*}$ is one in all entries, then $\Pr(Z_i = 1|\x_i) = \Pr(Z_k= 1|\x_k)$. 
\end{proposition}

 With this observation, we know that units matched together will have similar probabilities of being instrumented (in fact, as similar as possible, as finite data permits). %, within the available data. 
This will allow us to produce reliable estimates of $\lambda$ using our matched groups, provided that the data actually contain matches of sufficiently high quality. 

\subsection{FULL \airr\ PROBLEM}
\label{subsection:fullaer}
In the full version of the \air\ problem, the weights are chosen so that the variables used for each matched group have a useful quality: these variables together can create a high-quality predictive model for the outcomes. The weights become variable importance measures for each of the variables. 

%this match quality associated with each covariate is denoted by the weights, $\mathbf{w}$. The FLAME-IV algorithm works by computing an estimate of match quality for each covariate, $j$. \cut{by looking at how predictive of the outcome the covariate set is in absence of $j$.}
In order to determine the importance of each variable $j$, we use variable importance techniques to analyze machine learning models trained on a separate training set. Specifically, the units $1,\dots, n$ are divided into a training and a holdout set, the first is used to create matched groups and estimate causal quantities, and the second to learn the importance of each of the variables for match quality. Formally define the empirical predictive error on the training set, for set of variables $\vv$ as:
\begin{align*}
\widehat{\PE}_{\mathcal{F}}(\vv) &= \min_{f \in \mathcal{F}}\sum_{a \in \textrm{training}} (f(\vv \circ \x_{a}^{tr}, z_{a}^{tr}) - y_{a}^{tr})^2, 
\end{align*}
where $\mathcal{F}$ is some class of prediction functions. The empirical predictive error measures the usefulness of a set of variables. (The set of variables being evaluated are the ones highlighted by the indicator variables $\vv$.) 

We ensure that we always match using sets of variables $\vv$ that together have a low error $\widehat{\PE}_{\mathcal{F}}$. In fact, for each unit, if we cannot match on all the variables, we will aim to match on the set of variables for which the lowest possible prediction error is attained. Because of this, all matched groups are matched on a set of variables that together can predict outcomes sufficiently well.

The Full-\air\ problem can thus  be stated as: for all instrumented units $\unit$,
\begin{align*}
&\vv^{\unit*} \in \argmin\limits_{\vv \in \{0, 1\}^p}\widehat{\PE}_\mathcal{F}(\vv), \mbox{ such that:}\\
&\exists \;k \;\textit{ with }z_{k}=0 \textit{ and } \x_{k}\circ \vv^{\unit*} = \x_{\unit}\circ\vv^{\unit*},  
\end{align*}
 \textit{When importance weights are a linear function of the covariates, then solving the problem above is equivalent to solving the general \air\  problem}. An analogous result holds without IVs for the AME problem \citep{roy2017flame}. 
 %In addition, the simulations performed by \cite{roy2017flame} show that FLAME does a good job at selecting covariates with large values of $\mathbf{w}$ to match on even when this is not the case. 

In the standard Full-AME problem, there is no instrument, and each matched group must contain \textit{both treatment and control} units, whereas in the Full-\air\ case, the key is to match units so that instrumented units are matched with non-instrumented units \textit{regardless of treatment}. Intuitively, this makes sense because treatment uptake is in itself an outcome of instrumentation in the IV framework: a group with very large or very small numbers of treated or control units would imply that units with certain values of $\x$ are either highly likely or highly unlikely to respond to the instrument by taking up the treatment. %\textcolor{red}{Say something interesting here about matched groups can have the all treated or all control units and why that's allowed.}

\subsection{\FLAMEIVV: AN APPROXIMATE ALGORITHM FOR THE FULL-\airr\ PROBLEM}
%\textcolor{red}{The theory is cleaner without the balancing factor. Do not describe balancing factor until this section.}

We extend ideas from the Fast Large-scale Almost Matching Exactly (FLAME) algorithm introduced by \cite{roy2017flame} to approximately solve the \air\ problem. Our algorithm -- \FLAMEIV\ -- uses instrumental variables to create matched groups that have at least one instrumented and one non-instrumented unit within them. The procedure starts with an exact matching that finds all exact main matched groups. Then at each iteration \FLAMEIV\ iteratively chooses one covariate to drop, and creates matched groups on the remaining covariates. To decide which covariate to drop at each iteration, \FLAMEIV\ loops through the possibilities: it temporarily drops one covariate and computes the \textit{match quality} $\MQ$ after dropping this covariate. Then \FLAMEIV\ selects the covariate for which $\MQ$ was maximized during this loop. Match quality $\MQ$ is defined as a trade-off between prediction error, $\widehat{\PE}$ (which is defined in Section \ref{subsection:fullaer}) and a balancing factor, which is defined as:
{\scriptsize
\begin{align*}
    \BF &= \frac{\#\text{ matched non-instrumented}}{\#\text{ available non-instrumented}} 
    + \frac{\# \text{ matched instrumented}}{\# \text{ available instrumented}}
\end{align*}
}
$\MQ$ is computed on the holdout training dataset. In practice, the balancing factor improves the quality of matches by preventing \FLAMEIV\ from leaving too many units stranded without matched groups. That is, it could prevent all treated units from being matched to the same few control units when more balanced matched groups were possible. More details about the \FLAMEIV\ algorithm are in the supplement.
%because \red{from FLAME} the order in which covariates are dropped can affect the balance between instrumented and non-instrumented units. We would not want (for instance) to create a set of matched groups and then find at the next iteration, we had half the control group remaining and none of the treatment group. Ideally, we would match a large fraction of both treatment and control units at each iteration. FLAME considers the following balancing factor and among approximately equal choices of covariates to eliminate, chooses among those of lower balancing factor. 

%As in the \FLAME\ algorithm \cite{roy2017flame}, 
It is recommended to early-stop the algorithm before the $\MQ$\ drops by 5\% or more \citep{roy2017flame}. This way, the set of variables defining each matched group is sufficient to predict outcomes well (on the training set).
The details about early-stopping are in the supplement. 
\section{ESTIMATION}\label{sec:estimation}

Assuming that (A1) through (A5) and SUTVA hold, the LATE, $\lambda$, can be estimated in a consistent way \citep{imbens1994identification, angrist1996identification}; in this section we adapt common estimators for $\lambda$ to our matching framwork. Consider a collection of $m$ matched groups, $\MG_1, \dots, \MG_m$, each associated with a different value of $(\vv, \x)$. We estimate the average causal effect of the instrument on the treatment, $\ITT_{t, \ell}$ and on the outcome, $\ITT_{y, \ell}$, within each matched group, $\ell$, and then take the ratio of their weighted sums over all groups to estimate $\lambda$. 

We start with the canonical estimator for $\ITT_{y, \ell}$:
\begin{align}
    \widehat{\ITT}_{y,\ell} = \frac{\sum_{i \in \MG_\ell}y_{i}z_{i}}{\sum_{i \in \MG_\ell}z_{i}} - \frac{\sum_{i \in \MG_\ell}y_{i}(1-z_{i})}{\sum_{i \in \MG_\ell}(1-z_{i})}.\label{eq:itty}
\end{align}
Similarly, the estimator for the causal effect of the instrument on the treatment, $\ITT_{t, j}$, can be written as:
\begin{equation} 
    \widehat{\ITT}_{t, \ell} = \frac{\sum_{i \in \MG_\ell}t_iz_i}{\sum_{i \in \MG_\ell}z_i} - \frac{\sum_{i \in \MG_\ell}t_{i}(1-z_{i})}{\sum_{i \in \MG_\ell}(1-z_{i})}.\label{eq:ittt}
\end{equation}
From the form of $\lambda$ in Equation \eqref{eq:late} it is easy to see that, if the estimators in \eqref{eq:itty} and \eqref{eq:ittt} are unbiased for $\ITT_{y,\ell}$ and $\ITT_{t,\ell}$ respectively (which is true, for instance, when matches are made exactly for all units), then the ratio of their weighted average across all matched groups is a consistent estimator for $\lambda$:
\begin{equation}
    \hat{\lambda} = \frac{\sum_{\ell=1}^{m}n_\ell\widehat{\ITT}_{y,\ell}}{\sum_{\ell=1}^{m}n_\ell\widehat{\ITT}_{t,\ell}},\label{eq:lambdahat}
\end{equation}
where $n_\ell$ denotes the number of units in matched group $\ell$. A natural extension of this framework allows us to estimate the LATE within matched group $\ell$, defined as:
\begin{align}
\lambda_\ell &= \frac{1}{n_\ell}\sum\limits_{\substack{i \in \MG_\ell:\\t_{i\ell}(1) > t_{i\ell}(0)}} y_{i}(1) - y_{i}(0).
\end{align}
 This can be accomplished with the following estimator:
\begin{equation}
    \hat{\lambda}_{\ell} = \frac{\widehat{\ITT}_{y,\ell}}{\widehat{\ITT}_{t,\ell}}.\label{eq:lambdaihat}
\end{equation}
We quantify uncertainty around our estimates with asymptotic Confidence Intervals (CIs). To compute CIs for these estimators we adapt the approach laid out in \cite{imbens2015}. Details on variance estimators and computations are given in the supplement. 

In the following section, we present simulations that employ these estimators in conjunction with the algorithms presented in the previous section to estimate $\lambda$ and $\lambda_\ell$. The performance of our methodology is shown to surpass that of other existing approaches.

\section{SIMULATIONS}
We evaluate the performance of our method using simulated data. We compare our approach to several other methods including \textbf{two-stage least squares} \citep{angrist1991does,card1993using,wooldridge2010econometric}, and two other state-of-the-art nonparametric methods for instrumental variables, \textbf{full matching} \citep{kang2016full} and \textbf{nearfar matching} \citep{baiocchi2010building}. 
Full matching and nearfar matching find units that differ on the instrument while being close in covariate space according to a predefined distance metric. 
% minimize the distance between covariates while maximizing the difference in instrumented and non-instrumented units in each matched group. 
Both algorithms rely on a sample-rank Mahalanobis distance with an instrument propensity score caliper. 

We implement \FLAMEIVV\, using  \textit{bit-vector calculations}. More details about the implementation are in the supplementary materials.

In the first set of experiments, we compare the performance of the different methods on the estimation of local average treatment effects. In Experiment \ref{sec:groupexperiment} we demonstrate the power of \FLAMEIV\ for estimating individualized local average treatment effects. Experiment \ref{sec:timingexperiment} describes the scalability of the approach in terms of the number of covariates and number of units.
% In these experiments we study (1) how close matching estimates are to underlying ground truth treatment effects for the whole sample; (2) how close to ground truth are estimated treatment effects for individual units; (3) scalability of the algorithm.
%This distance metric is continuous and thus might perform poorly have on categorical data. 
% \subsection{Basic Simulations Settings}

Throughout, we generate instruments, covariates and continuous exposures based on the following structural equation model \citep{wooldridge2010econometric}:
\begin{align}
    T_i^\star &= k + \pi Z_i + \rho^T X_{i} + \xi_i\label{Eq:Tstar}
\end{align}
where $Z_i \sim \operatorname{Bernoulli}(0.5)$, and $\xi_i \sim N(0,0.8)$.  For important covariates, $X_{ij} \sim \operatorname{Bernoulli}(0.5)$. For unimportant covariates, $X_{ij} \sim \operatorname{Bernoulli}(0.1)$ in the control group, and $X_{ij} \sim \operatorname{Bernoulli}(0.9)$ in the treatment group. We discretize the exposure values $T_i^\star$ by defining:
\begin{align*}
T_i &= \ind_{[0.3 < T_i^\star \leq 0.6]} \\
    &+ 2 \times \ind_{[0.6 < T_i^\star \leq 1.0]} + 3 \times \ind_{[T_i^\star > 1.0]}.
\end{align*}
% Experiments were run on an Ubuntu 16.04.01 system with eight-core CPU  (Intel Core i7-4790 @ 3.6 GHz) and 8 GB RAM.
\subsection{ESTIMATION OF $\lambda$}\label{sec:populationexperiment}
% for the Entire Sample
In this experiment, outcomes are generated based on one of two homogeneous treatment effect models: a linear and a nonlinear model, respectively defined as:
\begin{eqnarray}
    Y_i &=&\sum_{j = 1}^{10}\alpha_j X_{ij} + 10T_i \label{eqn:homoeq}\\
    Y_i &=& \sum_{j = 1}^{10}\alpha_j X_{ij} + 10T_i + \sum_{1\leq j<\gamma\leq 5}X_{ij} X_{i\gamma}.\;\;\;\;\label{eqn:nonhomoeq}
\end{eqnarray}
Under both generation models, the true treatment effect is 10 for all individuals. There are 10 confounding covariates, 8 of which are important and 2 are unimportant. The importance of the variables is exponentially decaying with $\alpha_j =  0.5^j$. %(Variables with earlier indices are more important.)

We measure performance using the \textbf{absolute bias of the median}, i.e., the absolute value of the bias of the median estimate of 500 simulations and \textbf{median absolute deviation}, i.e., the median of the absolute deviations from the true effect, for each simulation. We present simulation results at varying levels of strength of the instrumental variable. This is measured by a concentration parameter, defined as the influence that the instrument has on treatment take-up.
This is represented by the concentration parameter $\pi$ in Eq. \eqref{Eq:Tstar}. Usually a concentration parameter below 10 suggests that instruments are weak  \citep{stock2002survey}.

%We present simulation results at different levels of \textit{concentration}. This is defined as the influence that the instrument has on treatment take-up, commonly known as \textit{strength} of the instrument. This is represented by the concentration parameter $\pi$ in Equation \eqref{Eq:Tstar}. Commonly, a concentration parameter below 10 suggests that the instruments are weak,  \citep{stock2002survey}.
%It is defined as the population value of the first-stage partial F-statistic for the instruments when regressing exposure $T$ on both covariates $X$ and instruments $Z$ \citep{bound1995problems}. A concentration parameter below 10 suggests that the instruments are weak \citep{stock2002survey}.
\cut{We use the concentration parameter here as a measure of strength values of instruments \citep{bound1995problems}, which is the population value of the first-stage partial F-statistic for the instruments by regressing exposures $T$ on both covariates $X$ and instruments $Z$. A concentration parameter below 10 suggests that the instruments are weak \citep{stock2002survey}.}
We also assess the performance of our methods by varying the size of training and holdout data. We generate two training and holdout datasets of different sizes: one with 1000 instrumented units and 1000 non-instrumented units, and one with 50 instrumented units and 50 non-instrumented units. For each case, we run each experiment 500 times for each of the algorithms.
\cut{In order to study this, we run each algorithm 500 times for each of the two models. In each experiment, we randomly generate the training and holdout datasets of different sizes, one with 1000 treatment units and one with 50 treatment units, both with equal numbers of control units. For each of the combinations of model and sample size, we aggregate 500 estimation results and measure the performance of each algorithm by computing \textbf{absolute bias of the median} (absolute value
of the bias of the median estimate) and \textbf{median absolute deviation}(median of the absolute deviations). }

\begin{figure}[h!]
\centering
\includegraphics[trim={1cm 0 3cm 0},clip, width=0.95\linewidth]{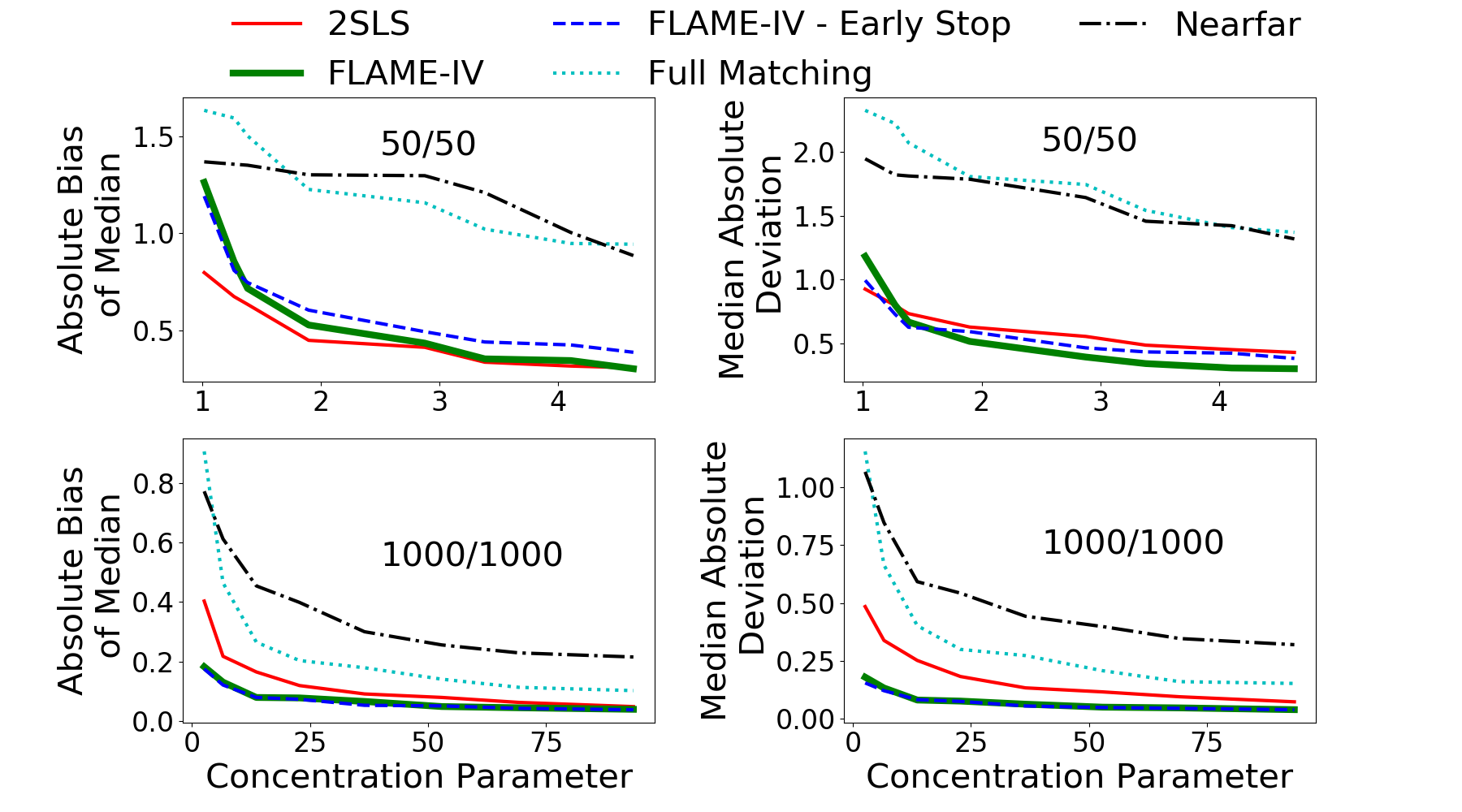}
\caption{Performance for linear generation model with various sample sizes. Here, 2SLS has an advantage because the data are generated according to a 2SLS model. \FLAMEIV\ (either early-stopping or run-until-no-more-matches) outperforms other methods on the large dataset, with smaller absolute bias of the median and median absolute deviation. On the smaller datasets, \FLAMEIV\ has a slightly larger bias than 2SLS but the smallest median absolute deviation among all  methods. 
\label{fig:pop_linear}}
\end{figure}

\begin{figure}[t]
\centering
\includegraphics[trim={1.4cm 0 3cm 0},clip, width=0.95\linewidth]{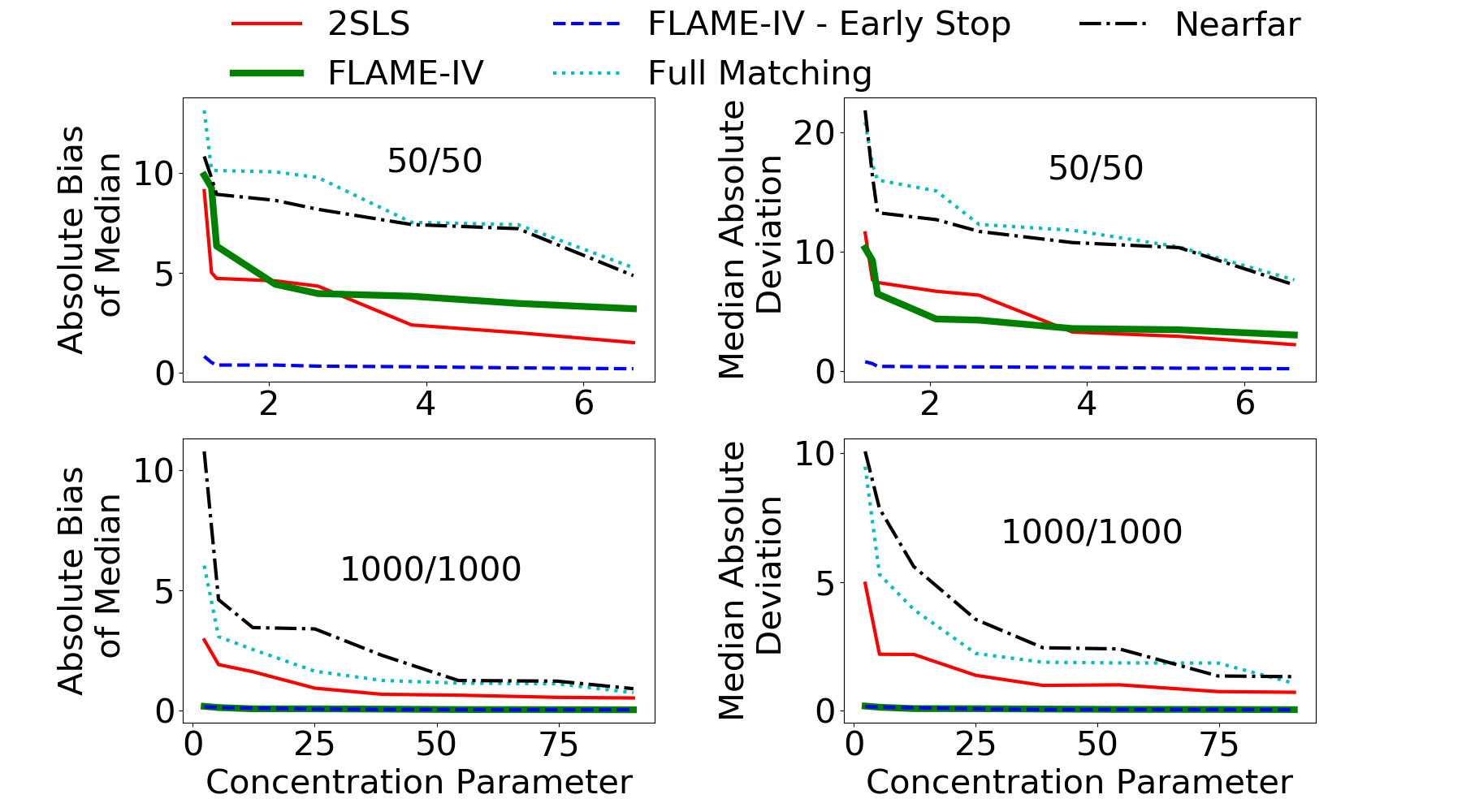}
\caption{Performance for nonlinear generation model with different sample sizes. Here, the 2SLS model is misspecified. \FLAMEIV\ (either early-stop or run-until-no-more-matches) outperforms other methods on both datasets, having smaller absolute bias of median and median absolute deviation. 
%\red{Note: Do I need to mention the smoothing method I used to smooth the lines? Which method did you use? If it was something simple, don't bother.
%I am using one-dimensional Gaussian filter with sigma equals to 2.5. Why are you doing this? Why don't you just plot the points?}
\label{fig:pop_nonlinear}}
\end{figure}

Figures~\ref{fig:pop_linear} and \ref{fig:pop_nonlinear} show the results of this experiment. All algorithms achieve better estimation accuracy when the instrument is stronger (i.e., more instrumented units take up the treatment). Figure~\ref{fig:pop_linear} shows results for the linear generation model, and Figure~\ref{fig:pop_nonlinear} shows results for the nonlinear generation model. As both figures show, \FLAMEIV\ with and without early-stopping generally outperform all other algorithms in terms of bias and deviation. This is likely because our methodology does not rely on a parametric outcome model and uses a discrete learned distance metric. The only exceptions are the left-upper plot on Figure \ref{fig:pop_linear} and Figure \ref{fig:pop_nonlinear}, which represents the bias results on small datasets (50 instrumented \& 50 noninstrumented). 2SLS has advantages here, because the amount of data is too small for powerful nonparametric methods like \FLAMEIV\ to fit reliably. \FLAMEIV's matching estimates lead to slightly larger bias than 2SLS. 
%This is expected and due to the outcome model for this simulation  being linear: 2SLS is a more efficient estimator in this case. In addition, \FLAMEIV\ only has few units to match, potentially leading to lower quality matches in this setting.

In the supplementary materials, we report results of similar experiments but with the additional inclusion of observed confounders of instrument assignment. We see no degradation in the performance. Result patterns with confounded instruments mimic those in Figures \ref{fig:pop_linear} and \ref{fig:pop_nonlinear}.

Next, we compare 95\% confidence intervals for each algorithm. The results are reported in Table ~\ref{table:simulation}. \FLAMEIV\ performs well on the nonlinear generation model, leading to the narrowest 95\% CI of all the methods. For the linear generation model, the 95\% CI for \FLAMEIV\ is narrower than the equivalent CIs for full matching and nearfar matching, but wider than 2SLS. Again, this is expected, and due to the correct parameterization of 2SLS with the linear generation model. More details about the confidence intervals are available in the supplement.

% Revised table: holdout = 15%, train = 85%, disjoint + early stopping with 5% stopping rule

\begin{table*}[t] %!htbp
\centering
\caption{Point Estimates for Linear and Nonlinear Models}
\resizebox{0.65\textwidth}{!}{%
\begin{threeparttable}
\begin{tabular}{lcccc}%{l@{\hskip 1.5in} c@{\hskip 0.275in} c@{\hskip 0.275in} c@{\hskip 0.275in} c@{\hskip 0.275in}}
% \hline
% \multicolumn{1}{ c }{} & \multicolumn{2}{ c }{ Vote Share} & \multicolumn{2}{ c }{ Voter Turnout}  \\
 \hline
  & \FLAMEIVV\ & 2SLS & Full-Matching & Nearfar Matching \\ [0.5ex] 
 \hline
 %\multicolumn{4}{ l }{ \textit{Panel A: Linear Model }}\\
 Linear Model & 10.15 & 10.16 & 10.96  & 11.23  \\
    & (9.72, 10.58) & (9.92, 10.40) & (10.14, 12.68) & (10.23, 12.89)  \\
    \midrule
 %\multicolumn{4}{ l }{ \textit{Panel B: Nonlinear Model}}\\
  Nonlinear Model  & 9.95  & 10.11 & 18.97 & 21.67   \\ [1ex] 
    & (9.47, 10.43) & (6.96, 13.25) & (11.35, 41.44) & (12.96, 45.71) \\
 \hline
\end{tabular}
\caption*{95\% confidence interval for each estimate is given in parentheses. The value of concentration parameter for linear model is 36.64, whereas the same for nonlinear model is 15.57.}%\textcolor{red}{I can't read this table - just center the Part A and Part B lines and it'd be fine.}} 
\end{threeparttable}}
\label{table:simulation}
% \vspace{-.5cm}
\end{table*}

\cut{
\begin{table}[]
\setlength{\tabcolsep}{5pt}
\centering
\caption{95\% Confidence Interval for Linear Model, Concentration Parameter = 36.64}
\small{
\begin{tabular}{|p{1.2cm}|p{1.5cm}|p{1.5cm}|p{1.5cm}|p{1.5cm}|} % @{} serves to suppress white space at ends of table
\toprule
   &  \multicolumn{3}{}{ Mean Squared Error (MSE) } \\ 
   & \hair & 2SLS  & Full Matching & Nearfar Matching\\ 
\cmidrule(l){1-4}
\midrule
Estimate     & 10.15 & 10.16 & 10.96 & 11.23\\
(95\% CI)      & (9.72, 10.58) & (9.92, 10.40) & (10.14, 12.68) & (10.23, 12.89) \\
Estimate     & 9.95 & 10.11 & 18.97 & 21.67 \\
(95\% CI)      & (9.47, 10.43) & (6.97, 13.25) & (11.35, 41.44) & (12.96, 45.71) \\
\bottomrule
\end{tabular}
\label{tab:mse_linear}
\end{table}
\begin{table}[]
\setlength{\tabcolsep}{5pt}
\centering
\caption{95\% Confidence Interval for Nonlinear Model, Concentration Parameter = 15.77}
\small{
\begin{tabular}{|p{1.2cm}|p{1.5cm}|p{1.5cm}|p{1.5cm}|p{1.5cm}|} % @{} serves to suppress white space at ends of table
\toprule
   &  \multicolumn{3}{}{ Mean Squared Error (MSE) } \\ 
   & \hair & 2SLS  & Full Matching & Nearfar Matching\\ 
\cmidrule(l){1-4}
\midrule
Estimation    & 9.95 & 10.11 & 18.97 & 21.67 \\
(95\% CI)    & (9.47, 10.43) & (6.97, 13.25) & (11.35, 41.44) & (12.96, 45.71) \\
\bottomrule
\end{tabular}
}

}
\label{tab:mse_nonlinear}
\end{table}
}

\subsection{ESTIMATION OF $\lambda_{\ell}$}\label{sec:groupexperiment}
\begin{figure}[h!]
\centering
\includegraphics[width=.45\textwidth]{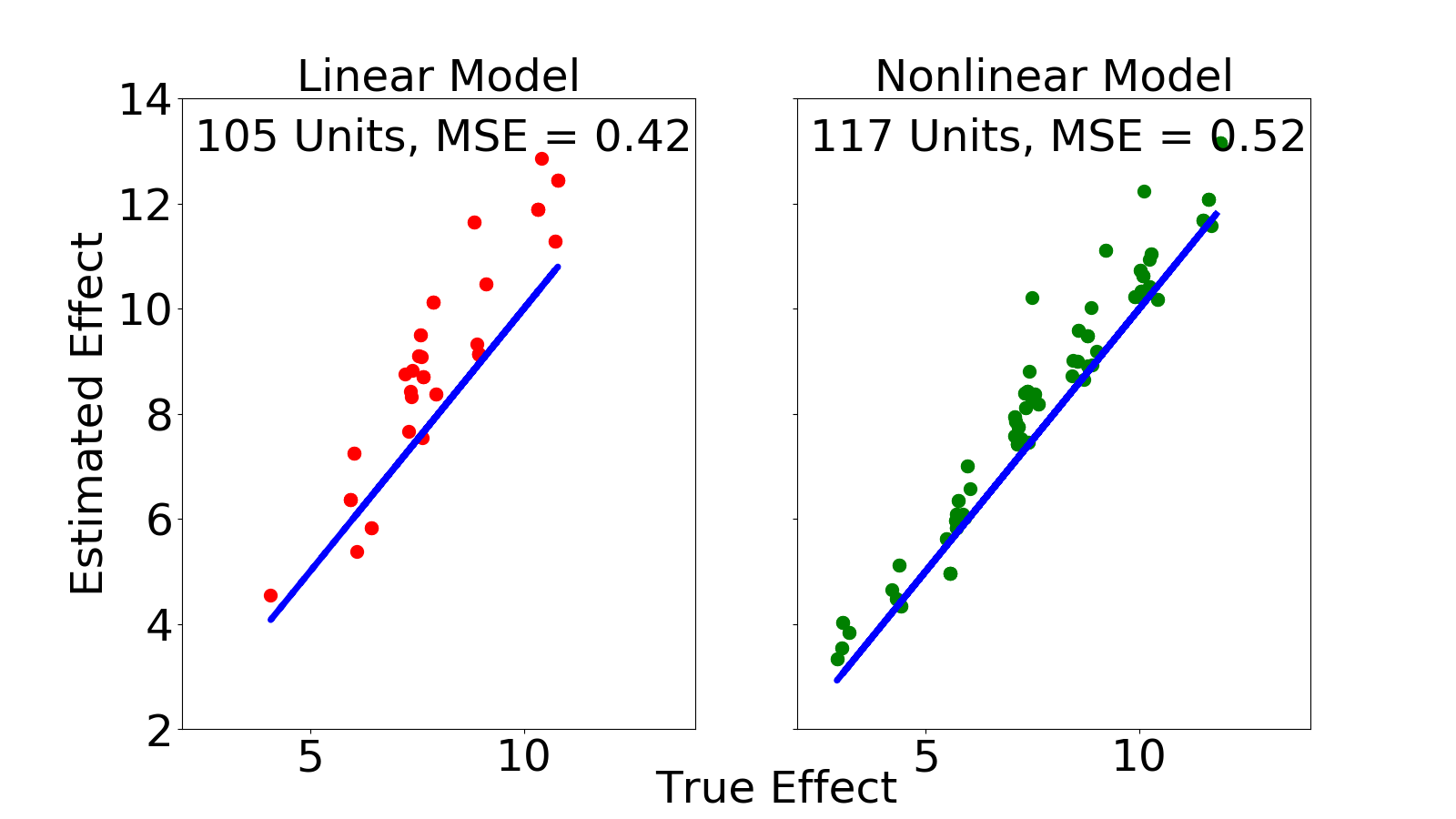}
\caption{True Individual Causal Effect vs$.$ Estimated Individual Causal Effect. The numbers on each plot represent the total number of instrumented units for calculating unit-level LATE, and MSE of our predictions. The concentration parameter is the same for the whole dataset, set to 288.84 for the linear outcome model, and 272.92 for the nonlinear outcome model.\label{fig:catt}}
\end{figure}

One advantage of the \air\ methodology is that it allows us to estimate LATE's on compliers (units for whom $t_i(1) > t_i(0)$) within each matched group. This results in more nuanced estimates of the LATE and in overall better descriptions of the estimated causal effects. We evaluate performance of \FLAMEIV\ in estimating matched group-level effects in a simulation study, with the estimators described in Section \ref{sec:estimation}.
To study how well \FLAMEIV\ estimates individual causal effects, we generate data with heterogeneous treatment effects. The new generation models, (\ref{eqn:heteroeq}) and (\ref{eqn:nonheteroeq}) below, are unlike the generation models in (\ref{eqn:homoeq}) and (\ref{eqn:nonhomoeq}), in that different individuals have different treatment effects. The two heterogeneous treatment effect data generation models are:
\begin{eqnarray}
\label{eqn:heteroeq}
    Y_i \hspace*{-5pt}&=& \hspace*{-5pt}\sum_{j = 1}^{T}\alpha_j X_{ij} + T_i\sum_{j = 1}^{10}\beta_j X_{ij}\\
    \label{eqn:nonheteroeq}
    Y_i \hspace*{-5pt}&=&  \hspace*{-5pt}\sum_{j = 1}^{T}\alpha_j X_{ij} + T_i\sum_{j = 1}^{10}\beta_j X_{ij} + \hspace*{-3pt}\sum_{\substack{ j = 1\dots5\\
    \gamma = 1\dots5 \\
    \gamma > j}}X_{ij} X_{i\gamma}.\;\;\;\;\;\;\;\;.
\end{eqnarray}
%Equation~\ref{eqn:heteroeq} is similar to Equation~\ref{eqn:homoeq}, but allows for heterogeneous treatment effects. 
Here $\alpha_{i} \sim N(10s,1)$ with $s \sim \operatorname{Uniform}\{-1,1\}$, $\beta_j \sim N(1.5,0.15)$. We generate 1000 treatment and 1000 control units from both models. We increased the value of the concentration parameter $\pi$ in Eq. (\ref{Eq:Tstar}) so that $Z$ has a strong effect on $T$ for the whole dataset. This is done to ensure appropriate treatment take-up within each group. Even with this adjustment, a few groups did not have any units take up treatment in the simulation. Results for these groups were not computed and are not reported in Figure \ref{fig:catt}. We estimate the LATE within each matched group ($\lambda_\ell$). Note that in groups where the instrument is very strong, the LATE will approximately equal the average treatment effect on the treated.

Experimental results for both data generation models are shown in Figure ~\ref{fig:catt}.  As we can see, our estimated effects almost align with true treatment effects and lead to relatively small estimation error for both linear and nonlinear generation models. Our algorithm performs slightly better when the generation model is linear.

\subsection{RUNNING TIME EVALUATION}\label{sec:timingexperiment}
\begin{figure}[h!]
\centering
\includegraphics[width=.45\textwidth]{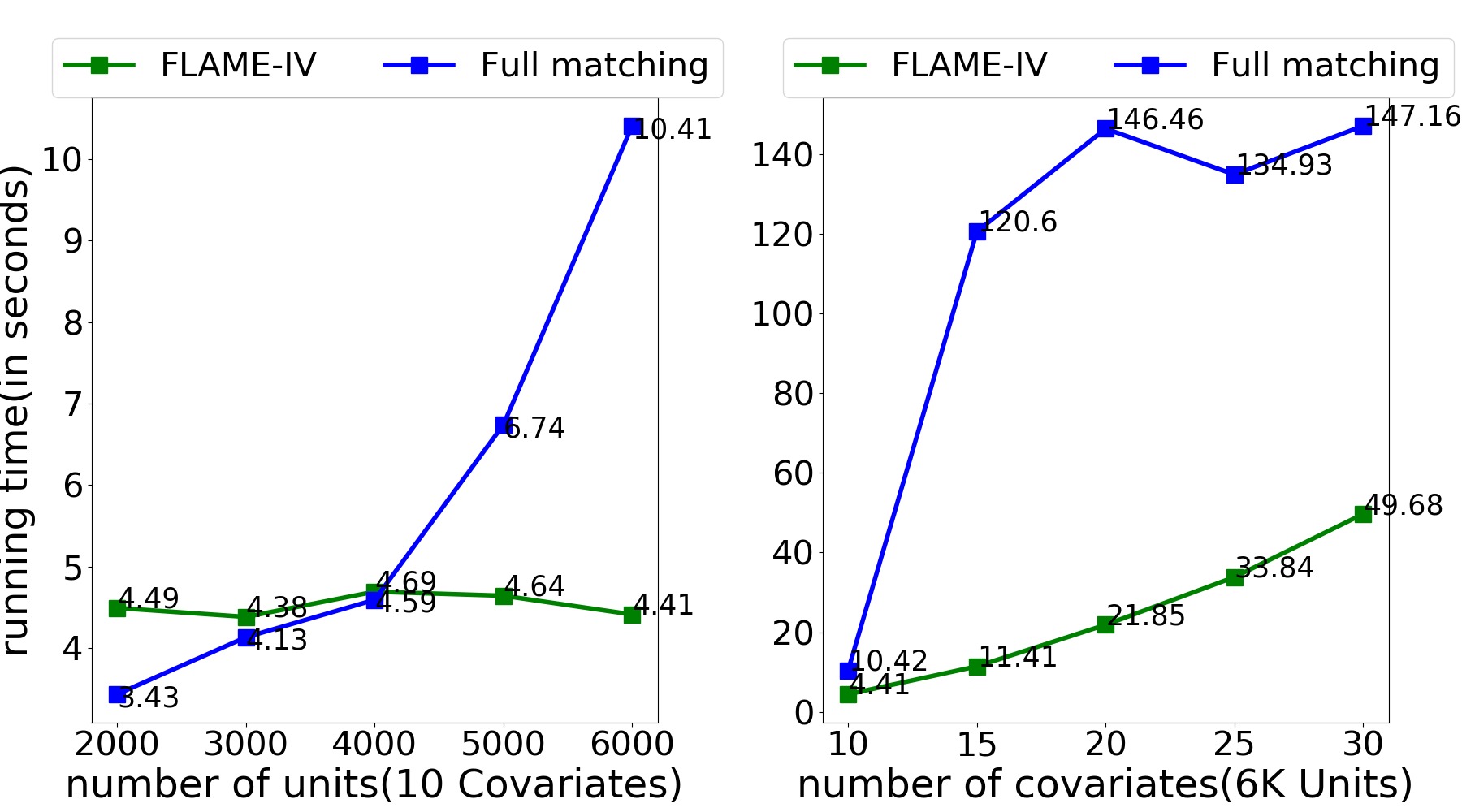}
\caption{Running Time for \FLAMEIVV\ and Full Matching. Left panel presents run time by varying number of units, whereas the right panel presents run time by varying number of covariates.  \label{fig:timing}} %\FLAMEIVV\ is more scalable due to its inherent design and bit-vector computations.}
\end{figure}

For the synthetic data generated by Section 5.2, Figure~\ref{fig:timing} compares the runtime of our algorithm against full matching. We computed the runtime by varying number of units (Figure \ref{fig:timing}, left panel) and by varying number of covariates (Figure \ref{fig:timing}, right panel). Each runtime is the average of five experiment results. The plot suggests that our algorithm scales well with both the number of units and number of covariates. Full matching depends on a Mahalanobis distance metric, which is costly to compute in terms of time. \FLAMEIV\ scales even better than full matching on a larger dataset with more units or covariates.  Experimental results about larger datasets are in the supplement. We note that the maximum number of units and covariates of full matching is also limited to the maximum size of vectors in R. Experiments were run on an Intel Core i7-4790 @ 3.6 GHz with 8 GB RAM and Ubuntu 16.04.01.

%\subsection{Sensitivity Analysis}

\section{WILL A FIVE-MINUTE DISCUSSION CHANGE YOUR MIND?}\label{sec:application}
In this section, we demonstrate the practical utility of our method by applying it to a real-world dataset. Since we do not observe the ground truth, we cannot evaluate the performance in terms of predictions, instead, we determine whether we can replicate the results of a previously published study. Specifically, we examine how door-to-door canvassing affects actual electoral outcomes; using experimental data generated by running a country-wide experiment during the 2012 French general election \citep{pons2018}. The original study estimates the effects of a door-to-door campaign in favor of François Hollande's \textit{Parti Socialiste} (PS) on two outcomes: voter turnout and share of votes for PS. The two outcomes are measured twice: once for each of the two rounds of voting that took place during the 2012 election. The units of analysis are geographically defined electoral precincts, often, but not always, comprised of different municipalities. 

% Revised table: holdout = 10%, train = 90%, disjoint
% Revised table: holdout = 15%, train = 85%, disjoint + early stopping with 5% stopping rule

\begin{table*}[t] %!htbp
\centering
\caption{Effect of Door-to-Door Canvassing on Electoral Outcomes}
%\begin{threeparttable}
\resizebox{0.8\textwidth}{!}{%
\begin{minipage}{\textwidth}
\begin{threeparttable}
\begin{tabular}{l@{\hskip 1.5in} c@{\hskip 0.275in} c@{\hskip 0.275in} c@{\hskip 0.275in} c@{\hskip 0.275in}}
 \hline
 \multicolumn{1}{ c }{} & \multicolumn{2}{ c }{ Vote Share} & \multicolumn{2}{ c }{ Voter Turnout}  \\
 \hline
  & First round & Second round & First round & Second round \\ [0.5ex] 
 \hline
 \multicolumn{4}{ l }{ \textit{Panel A: All Precincts}}\\
  & 0.02280 & 0.01593 & -0.00352 & -0.00634 \\ 
  & (0.00683) & (0.00827) & (0.00163) & (0.00158,) \\ 
%   \\ 
 \multicolumn{4}{ l }{ \textit{Panel B: Precincts by Income Levels}}\\
 Low & 0.02844 & 0.03903 & -0.00666  & -0.01505  \\
    & (0.00429) & (0.00562) & (0.00228) & (0.00254)  \\
 Medium & 0.01772  & 0.02090 & -0.00311 & -0.00070   \\
    & (0.00388) & (0.00434) & (0.00287) & (0.00333) \\
 High & 0.02560 & 0.04313 & -0.02717 &  -0.01367 \\
    & (0.02780) & (0.02752) & (0.01217) & (0.00538) \\
%  \\
 \multicolumn{4}{ l }{ \textit{Panel C: Precincts by Gender Majority}}\\
 Male & 0.05619  & -0.00442 & 0.00973 & -0.00056   \\ [1ex] 
    & (0.00879) & (0.00995) & (0.00376) & (0.00346) \\
 Female & 0.01640  & 0.00777  & -0.00692 & -0.00675 \\ [1ex] 
    & (0.00834) & (0.00719) & (0.00237) & (0.00239) \\
 \hline
\end{tabular}
\caption*{%The table reports causal effect of door-to-door canvassing on electoral outcomes for the two rounds of 2012  French  general  election. Two electoral outcomes are of interest, (1) vote share for Parti Socialiste (PS), and (2) voter turnout.
Columns 2 and 3 correspond to causal effects on vote share for PS, whereas Columns 4 and 5 reports causal effects on voter turnout. Panel A accounts for all the precincts and reports population causal effects. Panel B divides precincts by median income level and reports causal effect for each subgroup. Panel C divides precincts by gender-majority and reports associated causal effects. We use 15\% of the data as holdout training data and use a 5\% change in match quality as an early stopping rule.  Differences between our approach and the original paper's approach in estimated variances are mainly due to the strata used by the authors being marginally different from those produced with our methodology.
}
\end{threeparttable}
\end{minipage}}
%\end{threeparttable}
\label{table:causal}
\vspace{-0.5cm}
\end{table*}

The instrument in this case is pre-selection into campaign precincts: the 3,260 electoral precincts were clustered into strata of 5, among which 4 were randomly chosen and made available to conduct a campaign. The treatment is the decision of which of these four instrumented precincts to actually run campaigns in, as not all of the four instrumented precincts were actually chosen for door-to-door campaigns. The decision was based on the proportion of PS votes at the previous election within each precinct and the target number of registered citizens for each territory. These deciding factors evidently confound the causal relationship between treatment and outcomes. This setup provides an ideal setting for an Instrumental Variable design, where random pre-selection into campaign districts can be used to estimate the LATE of actual door-to-door campaigns on both turnout and PS vote share. 
 
We replicate the original study's results by running our algorithm on the data without explicitly accounting for the strata defined by the original experiment. Since some of the covariates used for matching are continuous, we coarsen them into 5 ordinal categories. We coarsen turnout at the previous election and PS vote share at the previous election into 10 categories instead, as these variables are particularly important for matching and we would like to make more granular matches on them. Results from applying our methods to the data from the study are presented in Table \ref{table:causal}. Columns 2 and 3 shows results for PS vote share as an outcome, and the last two columns for voter turnout as an outcome. Results are presented disaggregated by each round of election. %and as an average of both rounds.

\textit{Panel A} provides LATE estimates from \FLAMEIV. Unlike the earlier study \citep{pons2018}, our estimates are independent of the strong parametric assumptions of 2SLS. We reach conclusions similar to those of the original paper, finding no positive effect of canvassing on voter turnout and a positive statistically significant effect on vote share for PS. In general, our standard error estimates are similar to those obtained with 2SLS, however more conservative due to the non-parametric nature of the estimators we employ. Interestingly, our estimate of the effect of canvassing on vote share has a \textit{greater} magnitude than the original analysis, while our estimate for the effect of canvassing on voter turnout is nearly the same as the original paper's. 

Our methodology also allows an improvement on the original analysis by estimating effects of door-to-door campaigns on the two outcomes for particular subgroups of interest. LATE estimates for income and gender subgroups are reported in \textit{Panel B} and \textit{Panel C} of Table \ref{table:causal}. The income subgroups are defined by median income, whereas gender subgroups are defined by share of female population in each precinct. We find that canvassing was more effective in increasing the vote share for PS, in the first round of the election, in precincts where male population is in the majority. We also find that canvassing had negative effect on voter turnout in low income precincts, but positive effect on voter share for PS. The combination of these results show that canvassing was successful in convincing voters to switch their votes in favour of François Hollande.

In the supplement we show two example matched groups output by \FLAMEIV. In this case the algorithm was successful in separating localities with low support for PS from localities in which support for PS was greater. These examples highlight how the algorithm can produce meaningful and interpretable groups, while reducing potential for confounding by observed covariates. 

In conclusion, the results of our analysis of the voter turnout data clearly show that our method produces novel and interesting results when applied to real-world scenarios, independently of strong parametric assumptions, and with a simple interpretable framework.
\section{CONCLUSION}

Matching methods can be extremely powerful: they are both highly nonparametric and interpretable to users, allowing them to trust and troubleshoot their models more easily. Our approach to matching for  instrumental variables accounts for the limitations faced by existing methods. We improve on 2SLS by using a highly non-parametric powerful modeling approach. We retain interpretability unlike traditional machine learning approaches by using matching. We improve on existing matching methods by \textit{learning} an interpretable distance metric on a training set. Our methodology also provides a systematic way to account for nuisance variables, and to achieve consistently high quality matching outcomes. The algorithm can be implemented easily within most common database systems for optimal performance. It scales well to large datasets. It achieves a balance between interpretability, scalability, trustworthiness, and modeling power that is unsurpassed by any other method for IV analysis. Code is publicly available at:
\url{https://github.com/almost-matching-exactly}
% \textcolor{red}{url removed to preserve blind review}.

\subsubsection*{Acknowledgements} 
This work was supported in part by NIH award
1R01EB025021-01, NSF awards IIS-1552538 and IIS1703431, a DARPA award under the L2M program, and a Duke University Energy Initiative Energy Research Seed Fund  (ERSF) grant.

\balance
%\section*{References}
\newpage
\bibliographystyle{plainnat}
\renewcommand{\bibname}{\subsubsection*{References}}
\bibliography{biblio}

\clearpage
\section{Supplement}
\subsection{Proof of Lemma \ref{Theorem:Weighted Hamming Distance}}
Since the result is exactly symmetric when non-instrumented units are matched we prove it only for the case when instrumented units are matched. Assume $\ww \in \mathbb{R}^p$. For a given unit $i$ with $z_i$ = 1, suppose we could find a $\vv^{\unit*}$ as defined in the AME-IV problem. Let us define another unit $k$ with $z_k$ = 0, and $\x_{k}\circ \vv^{\unit*} = \x_{\unit}\circ\vv^{\unit*}$, by definition of $\MG(\vv^{\unit*},\x_i)$ it must be that $\x_k \in \MG(\vv^{\unit*},\x_i)$. So $\one_{[\x_i \neq \x_k]} = \mathnormal{J} - \vv^{\unit*}$, where $\mathnormal{J}$ is a vector of length $p$ that has all entries equals to 1.

Assume there is another unit $j$ with $z_j$ = 0, and $j \neq k$. 

If $j \in \MG(\vv^{\unit*},\x_i)$, then $\one_{[\x_i \neq \x_j]} = \mathnormal{J} - \vv^{\unit*}$. So 
\begin{align*}
\ww^T\one_{[\x_i \neq \x_k]} = \ww^T(\mathnormal{J} - \vv^{\unit*}) = \ww^T\one_{[\x_i \neq \x_j]}
\end{align*}

If $j \notin \MG(\vv^{\unit*},\x_i)$, let us define $\vv^{\unit{j}} = \mathnormal{J} - \one_{[\x_\unit \neq \x_j]}$, obviously $\vv^{\unit{j}} \neq \vv^{\unit*}$. Since $\vv^{\unit*} \in \argmax\limits_{\vv\in\{0,1\}^p} \vv^T\mathbf{w}$, we have:
\begin{align*}
    \ww^T\one_{[\x_i \neq \x_k]} &= \ww^T(\mathnormal{J} - \vv^{\unit*})\\
                          &= \ww^T - \ww^T\vv^{\unit*}\\
                          &< \ww^T - \ww^T\vv^{\unit{j}}\\
                          &= \ww^T(\mathnormal{J} - \vv^{\unit{j}})\\
                          &= \ww^T\one_{[\x_i \neq \x_j]}.
\end{align*}
Therefore,
\begin{equation*}
k \in \argmin\limits_{\substack{j = 1, \dots, n\\Z_j = 0}}\ww^T\one_{[\x_i \neq \x_j]}.
\end{equation*}
This concludes the proof.

\subsection{Asymptotic Variance and Confidence Intervals for LATE Estimates}
To construct estimators for the  variance of $\hat{\lambda}$ we use an asymptotic approximation, that is, we will try to estimate the asymptotic variance of $\hat{\lambda}$, rather than its small sample variance. The strategy we use to do this is the same as \cite{imbens2015}, with the difference that our data is grouped: we adapt their estimators to grouped data using canonical methods for stratified sampling. In order to define asymptotic quantities for our estimators, we must marginally expand the definitions of potential outcomes introduced in our paper. In practice, while our framework has been presented under the assumption that the potential outcomes and treatments are fixed, we now relax that assumption and instead treat $y_i(1), y_i(0), t_i(1), t_i(0)$ as realizations of random variables $Y_i(1), Y_i(0), T_i(1), T_i(0)$, which are drawn from some unknown distribution $f(Y_i(1), Y_i(0), T_i(1), T_i(0))$. In this case the SUTVA assumption requires that each set of potential outcomes and treatments is independently drawn from the same distribution for all units. As usual, lowercase versions of the symbols above denote observed realizations of the respective random variables. 

The asymptotic behaviour of our method is straightforward. Since the covariates we consider are discrete (say binary for convenience) there are only a finite number of possible covariate combinations one can observe. If the sample size $n$ increases and the probability of observing all combinations of covariates is positive then asymptotically all possible combinations of covariates will be observed. In fact, most units will be matched exactly when $n \gg p$. This means that our matched groups will only contain exactly matched units, and therefore be exactly equal to a stratified fully randomized experiment in which the strata are the matched groups, by our Assumption 3 of our paper. By this principle, asymptotic results for IV estimation in stratified experiments, such as those in \cite{imbens2015}, apply asymptotically.

Recall as well that in this scenario we have a set of $m$ matched groups $\MG_1, \dots \MG_m$ indexed by $\ell$, such that each unit is only in one matched group. We denote  the number of units in matched group $\ell$ that have $z_i = 1$ with $n_\ell^1$ and the number of units in matched group $\ell$  with $z_i = 0$ with $n_\ell^0$. Finally the total number of units in matched group $\ell$ is $n_\ell = n_\ell^0 + n_\ell^1$.

We make all the assumptions listed in Section\ref{methodology} but we must require a variant of (A3), to be used instead of it. This assumption is:  \\
\textbf{(A3')} $\Pr(Z_i = 1| i \in \MG_\ell) = \Pr(Z_k = 1|k \in \MG_\ell) = \frac{n_\ell^1}{n_\ell}, \forall i, k$.\\
That is, if two units are in the same matched group, then they have the same probability of receiving the instrument. This probability will be equal to the ratio of instrument 1 units to all units in the matched group because we hold these quantities fixed. Note that this more stringent assumption holds when matches are made exactly, and is common in variance computation  for matching estimators (see, for example, \cite{kang2016full}). 

We keep our exposition concise and we do not give explicit definitions for our variance estimands. These are all standard and can be found in \cite{imbens2015}.

We have to start from estimating variances of observed potential outcomes and treatments within each matched group. We do so with the canonical approach: %\textcolor{red}{make all non-r.v.'s lower case?}
\begin{align*}
\hat{s}^2_{\ell0} &= \frac{1}{n_\ell^0 - 1}\sum_{i \in \MG_\ell} \left(y_i(1 - z_i) - \frac{1}{n_\ell^0}\sum_{i \in \MG_\ell}y_i(1 - z_i)\right)^2\\
\hat{s}^2_{\ell1} &= \frac{1}{n_\ell^1 - 1}\sum_{i \in \MG_\ell} \left(y_iz_i - \frac{1}{n_\ell^1}\sum_{i \in \MG_\ell}y_iz_i\right)^2\\
\hat{r}^2_{\ell0} &= \frac{1}{n_\ell^0 - 1}\sum_{i \in \MG_\ell} \left(t_i(1 - z_i) - \frac{1}{n_\ell^0}\sum_{i \in \MG_\ell}t_i(1 - z_i)\right)^2\\
&= 0\\
\hat{r}^2_{\ell1} &= \frac{1}{n_\ell^1 - 1}\sum_{i \in \MG_\ell} \left(t_iz_i - \frac{1}{n_\ell^1}\sum_{i \in \MG_\ell}t_iz_i\right)^2, 
\end{align*}
where: $\hat{s}^2_{\ell0}$ is an estimator for the variance of potential responses for the units with instrument value 0 in matched group $\ell$, $\hat{s}^2_{\ell1}$ for the variance of potential responses for the units with instrument value 1 in matched group $\ell$, $\hat{r}^2_{\ell0}$ for the variance of potential treatments the units with instrument value 0 in matched group $\ell$, and $\hat{r}^2_{\ell1}$ is an estimator for the variance of potential treatments for the units with instrument value 1 in matched group $\ell$. The fact that $\hat{r}_{\ell0}^2 = 0$ follows from Assumption A4. 

We now move to variance estimation for the two $\ITT$s. Conservatively biased estimators for these quantities are given in \cite{imbens2015}. These estimators are commonly used in practice and simple to compute, hence why they are often preferred to unbiased but more complex alternative. We repeat them below:
\begin{align*}
\widehat{Var}(\widehat{\ITT}_{y}) &= \sum_{\ell = 1}^m \left(\frac{n_\ell}{n}\right)^2\left(\frac{\hat{s}_{\ell1}^2}{n_\ell^1} + \frac{\hat{s}_{\ell0}^2}{n_\ell^0}\right)\\
\widehat{Var}(\widehat{\ITT}_{t}) &= \sum_{\ell = 1}^m \left(\frac{n_\ell}{n}\right)^2\frac{\hat{r}_{\ell1}^2}{n_\ell^1}.
\end{align*}

To estimate the asymptotic variance of $\hat{\lambda}$ we also need estimators for the covariance of the two $\ITT$s both within each matched group, and in the whole sample. Starting with the former, we can use the standard sample covariance estimator for $Cov(\widehat{\ITT}_{y\ell}, \widehat{\ITT}_{t\ell})$:
\begin{align*}
\widehat{Cov}(\widehat{\ITT}_{y\ell}, \widehat{\ITT}_{t\ell}) &= \frac{1}{n_\ell^1(n_\ell^1 - 1)}\\
&\times\sum_{i \in \MG_\ell} \left(y_iz_i - \frac{1}{n_\ell^1}\sum_{i \in \MG_\ell} y_iz_i\right)\\
&\qquad\,\times\left(t_iz_i - \frac{1}{n_\ell^1}\sum_{i \in \MG_\ell} t_iz_i\right). 
\end{align*}
The reasoning behind why we use only units with instrument value 1 to estimate this covariance is given in \cite{imbens2015}, and follows from A4. We can use standard techniques for covariance estimation in grouped data to combine the estimators above into an overall estimator for $Cov(\widehat{\ITT}_y, \widehat{\ITT}_t)$ as follows:
\begin{align*}
\widehat{Cov}(\widehat{\ITT}_y, \widehat{\ITT}_t) &= \sum_{\ell = 1}^m\left(\frac{n_\ell}{n}\right)^2\widehat{Cov}(\widehat{\ITT}_{y\ell}, \widehat{\ITT}_{t\ell}). 
\end{align*}
Once all these estimators are defined, we can use them to get an estimate of the asymptotic variance of $\hat{\lambda}$. This quantity is obtained in \cite{imbens2015} with an application of the delta method to convergence of the two $\ITT$s. The final estimator for the asymptotic variance of $\hat{\lambda}$, which we denote by $\sigma^2$, is given by:
\begin{align*}
\hat{\sigma}^2 &= \frac{1}{\widehat{\ITT}_t^2}\widehat{Var}(\widehat{\ITT}_y) + \frac{\widehat{\ITT}_y^2}{\widehat{\ITT}_t^4}\widehat{Var}(\widehat{\ITT}_t)  \\
&- 2\frac{\widehat{\ITT}_y}{\widehat{\ITT}_t^3}\widehat{Cov}(\widehat{\ITT}_y, \widehat{\ITT}_t). 
\end{align*}
Using this variance, $1 - \alpha$\% asymptotic confidence intervals can be computed in the standard way. 

\subsection{The FLAME-IV Algorithm}\label{sec:algo-generic}
We adapt the Algorithm described in \cite{roy2017flame} to the IV setting. The algorithm is ran as described in that paper, except instrument indicator is used instead of the treatment indicator as input to the algorithm. Here we give a short summary of how the algorithm works and refer to \cite{roy2017flame} for an in-depth description.

FLAME-IV takes as inputs a training dataset $D = \{(x_i, t_i, z_i, y_i)\}_{i=1}^n$, consisting of covariates, instrument indicator, treatment indicator, and outcome for every unit that we wish to match on, as well as a holdout set $D^H$ consisting of the same variables for a different set of units that aren't used for matching but to evaluate prediction error and match quality. The algorithm then first checks if any units can be matched exactly with at least one unit with the opposite instrument indicator. If yes, all the units that match exactly are put into their own matched group and removed from the pool of units to be matched. After this initial check, the algorithm starts iterating through the matching covariates: at each iteration, Match Quality is evaluated on the holdout set after removing each covariate from the set of matching covariates. The covariate whose removal leads to the smallest reduction in \MQ, is discarded and the algorithm proceeds to look for exact matches on all the remaining covariates. Units that can be matched exactly on the remaining covariates are put into matched groups, and removed from the set of units to be matched. Note that \MQ is recomputed after removing each remaining covariate at each iteration because the subset of covariates that it is evaluated on always is always smaller after each iteration (it does not include the covariate removed prior to this iteration). The algorithm will proceed in this way, removing covariates one by one, until either: a) \MQ\ goes below a pre-defined threshold, b) all remaining units are matched, or c) all covariates are removed. Experimental evidence presented in \cite{roy2017flame} suggests a threshold of 5\% of the prediction error with all of the covariates.  The matched groups produced by the algorithm can then be used with the estimators described in the paper to estimate desired treatment effects. Units left unmatched after the algorithm stops are not used for estimation. The algorithm ensures that at least one instrumented and one non-instrumented unit are present in each matched groups, but has no guarantees on treatment and control units: matched groups that do not contain either treated or control units are not used for estimation. 

One of the strengths of FLAME-IV is that it can be implemented in several ways that guarantee performance on large datasets. An implementation leveraging bit vectors is described in \cite{roy2017flame}, optimizing speed when datasets are not too large. A native implementation of the algorithm on any database management system that uses SQL as a query language is also given in the same paper: this implementation is ideal for large relational databases as it does not require data to be exported from the database for matching. 

While FLAME-IV is a greedy solution to the AME-IV problem, an optimal solution could be obtained by adapting the DAME (Dynamic Almost Matching Exactly) procedure described in \cite{dieng2018collapsing} to the IV setting by using instrument indicators as treatment indicators in the input to the algorithm. Resulting matched groups with no treated or control units should be discarded as we do here. The same estimators we employ in this paper can also be employed with the same properties for matched groups constructed with this methodology.

\begin{figure}[t]
\centering
\includegraphics[width=.45\textwidth]{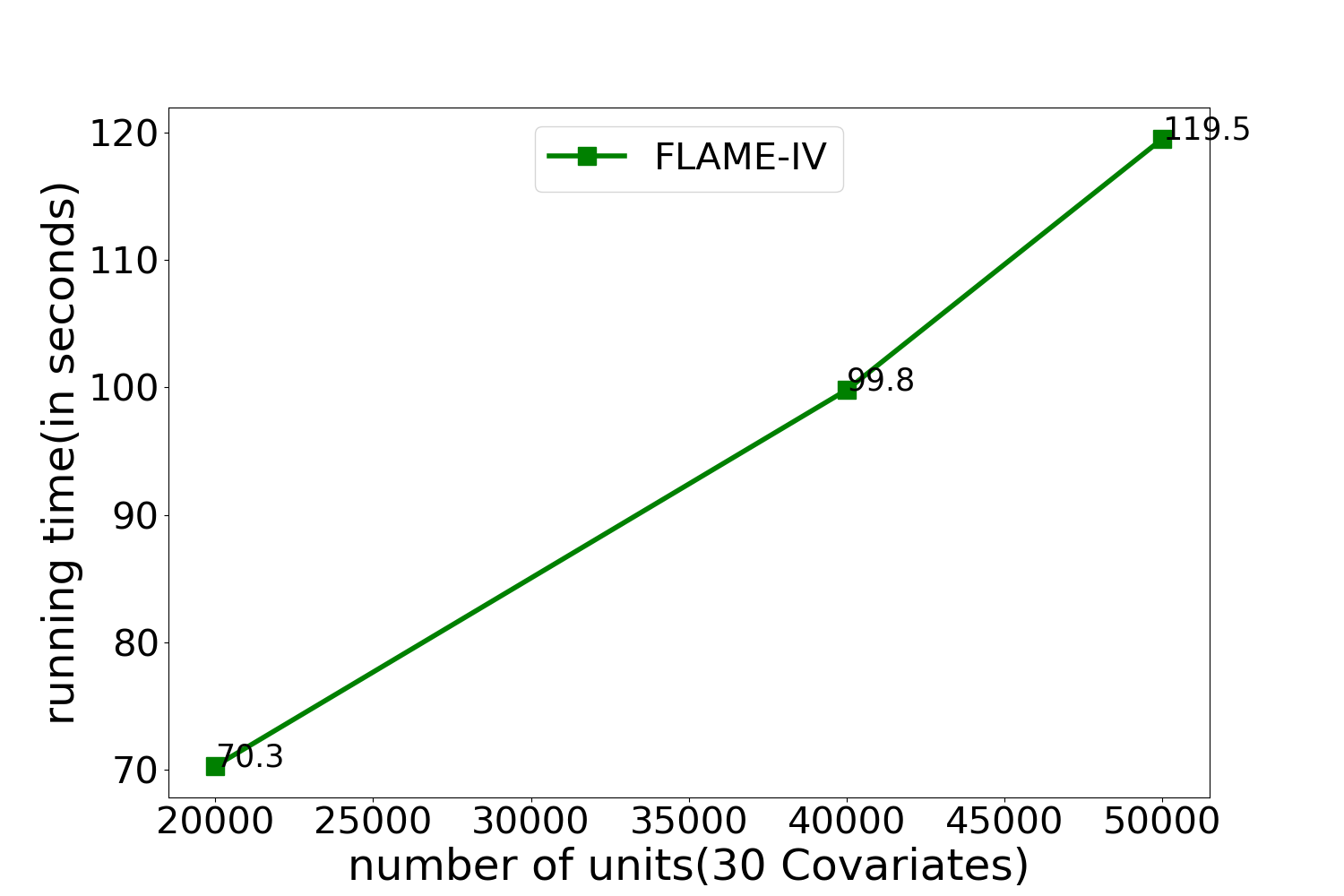}
\caption{Running Time for \FLAMEIVV\ on large dataset. \label{fig:timing_large}}
\end{figure}

\subsection{More Running Time Results on Large Dataset}
Figure ~\ref{fig:timing_large} shows the results of running time for \FLAMEIVV\ on a larger dataset. The running time is still very short($<$ 2 min) on the large dataset for \FLAMEIVV. Full matching can not handle a dataset of this size.

\subsection{Additional Simulations with Confounded Instrument Assignment}
Here we present results from simulations similar to those in Section  \ref{sec:populationexperiment}, but where, in addition to treatment assignment, instrument assignment is also confounded. 
Instrument is assigned as follows:
\begin{eqnarray}
\label{eqn:confound_iv}
    Z_i^{\prime} = \rho X_i^{\prime} \\
    Z^* = \Median\{Z_i^{\prime}\} \forall i \\
    Z_i = \ind_{[Z_i^{\prime} \geq Z^*]}
\end{eqnarray}
where $X_i^{\prime}$ contains last two variables for unit $i$, and $\rho \sim N(0.1,0.01)$. 

\begin{figure}
    \centering
    \includegraphics[width=.45\textwidth]{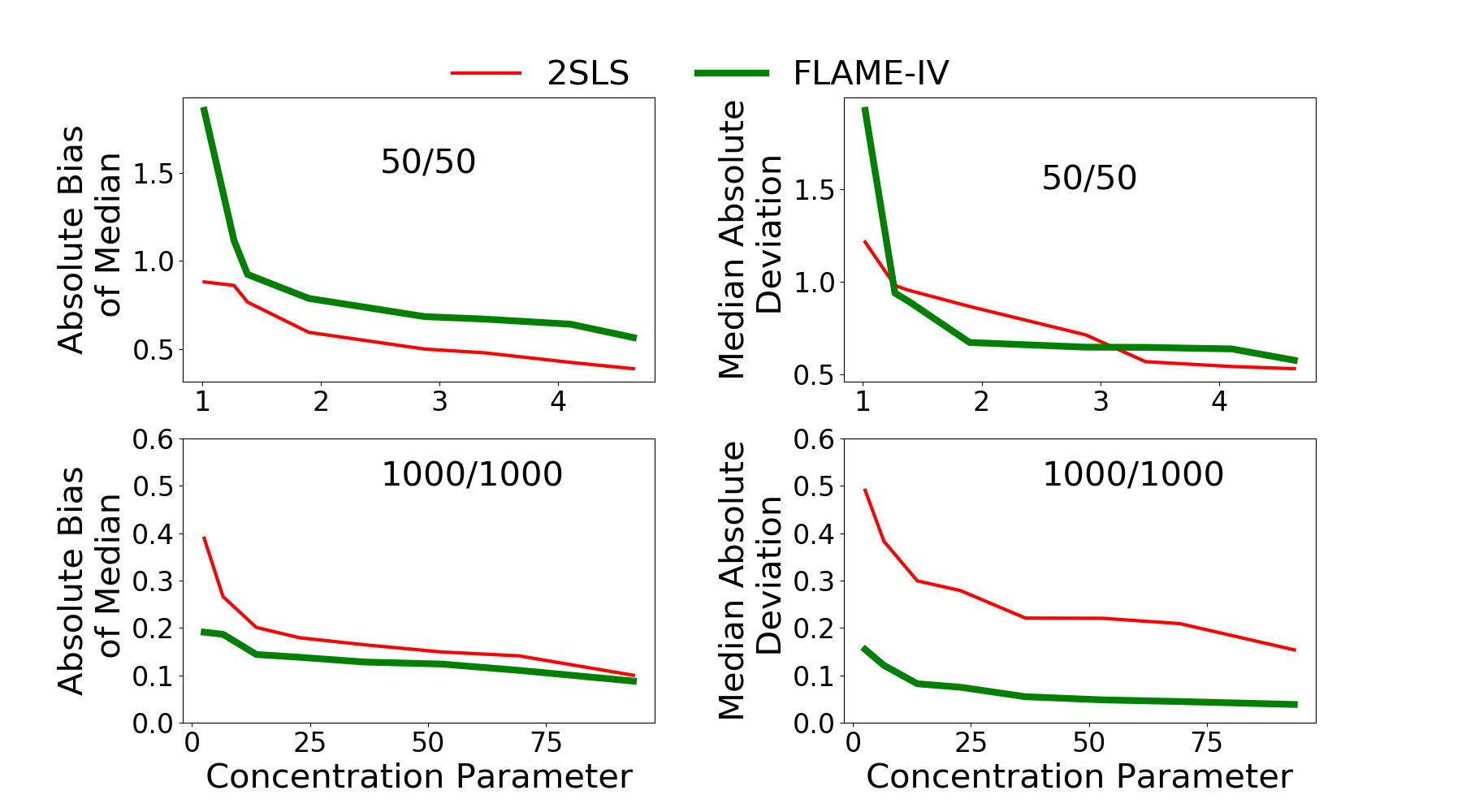}
    \caption{Performance for linear generation model with confounded instrument at various sample sizes. Here, 2SLS has an advantage because the data are generated according to a 2SLS model. \FLAMEIV\ (either early-stopping or run-until-no-more-matches) performs similarly to 2SLS on large datasets, with smaller absolute bias of the median and median absolute deviation. On the smaller datasets, \FLAMEIV\ has a slightly larger bias than 2SLS.}
    \label{Fig:instrument_linear_conf}
    \includegraphics[width=.45\textwidth]{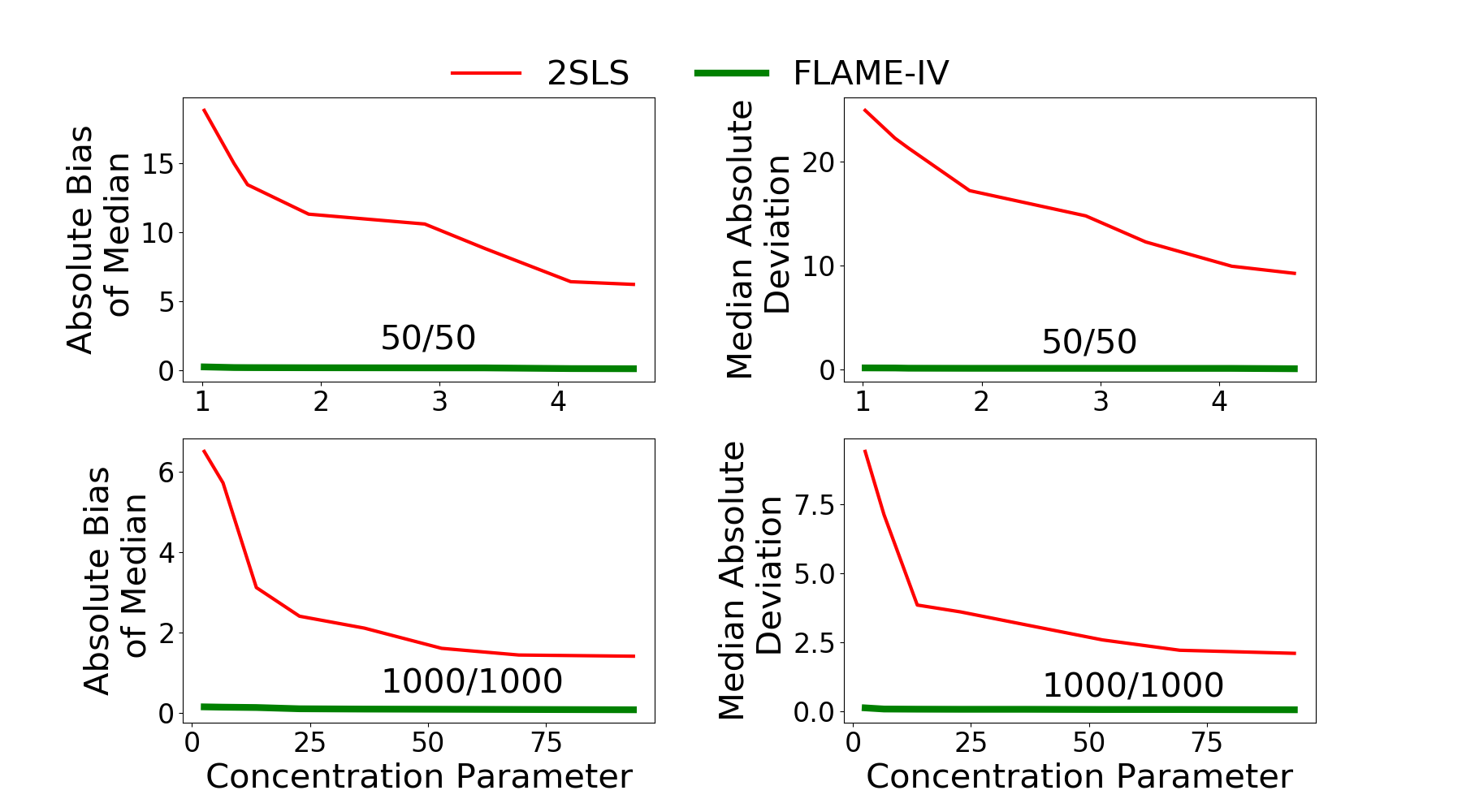}
    \caption{Performance for nonlinear generation model with confounded instrument assignment and different sample sizes. Here, the 2SLS model is misspecified. \FLAMEIV\ (either early-stop or run-until-no-more-matches) outperforms  2SLS on both datasets, having smaller absolute bias of median and median absolute deviation. }
    \label{Fig:instrument_nonlinear_conf}
\end{figure}

Results for a linear outcome model, same as in Equation \eqref{eqn:homoeq} are displayed in Figure \ref{Fig:instrument_linear_conf}, and results for a nonlinear outcome model as in Eq: \eqref{eqn:nonhomoeq} are displayed in Figure \ref{Fig:instrument_nonlinear_conf}. Results are largely similar to those obtained when instrument assignment is unconfounded. This suggests that our method performs equally well when instrument assignment is confounded. 

%\balance
%\begin{center}
\begin{table*}[t]
    \centering
\resizebox{\textwidth}{!}{%
\begin{tabular}{llllllrr}
\toprule
Territory & Last Election  & Last Election  &  Population & Share Male & Share  &  Treated &  Instrumented \\
& PS Vote Share & Turnout & (in thousands) & & Unemployed & & \\
\midrule
Matched Group 1\\
\midrule
Plouguenast et environs &         (0.01, 0.05] &           (0.77, 0.88] &    (0, 450] &  (0.47, 0.57] &   (0, 0.1] &        0 &        1 \\
Lorrez-le-Bocage-Préaux et environs &         (0.01, 0.05] &           (0.77, 0.88] &    (0, 450] &  (0.47, 0.57] &   (0, 0.1] &        0 &        1 \\
La Ferté-Macé et environs &         (0.01, 0.05] &           (0.77, 0.88] &    (0, 450] &  (0.47, 0.57] &   (0, 0.1] &        0 &        1 \\
Mundolsheim et environs &         (0.01, 0.05] &           (0.77, 0.88] &    (0, 450] &  (0.47, 0.57] &   (0, 0.1] &        1 &        1 \\
Paris, 7e arrondissement &         (0.01, 0.05] &           (0.77, 0.88] &  (1,800, 2,250] &  (0.47, 0.57] &       (0.1, 0.2] &        0 &        1 \\
Sainte-Geneviève et environs &         (0.01, 0.05] &           (0.77, 0.88] &    (0, 450] &  (0.47, 0.57] &   (0, 0.1] &        0 &        0 \\
Cranves-Sales et environs &         (0.01, 0.05] &           (0.77, 0.88] &    (0, 450] &  (0.47, 0.57] &   (0, 0.1] &        0 &        0 \\
Hem et environs &         (0.01, 0.05] &           (0.77, 0.88] &    (0, 450] &  (0.47, 0.57] &   (0, 0.1] &        0 &        1 \\
Legé et environs &         (0.01, 0.05] &           (0.77, 0.88] &    (0, 450] &  (0.47, 0.57] &   (0, 0.1] &        0 &        1 \\
Moûtiers et environs &         (0.01, 0.05] &           (0.77, 0.88] &    (0, 450] &  (0.47, 0.57] &   (0, 0.1] &        0 &        0 \\
Paris, 7e arrondissement &         (0.01, 0.05] &           (0.77, 0.88] &  (1,800, 2,250] &  (0.47, 0.57] &       (0.1, 0.2] &        0 &        1 \\
Craponne-sur-Arzon et environs &         (0.01, 0.05] &           (0.77, 0.88] &    (0, 450] &  (0.47, 0.57] &   (0, 0.1] &        0 &        0 \\
\midrule
Matched Group 2\\
\midrule
Nantes &(0.19, 0.22] &  (0.66, 0.77] &  (0, 450] &  (0.47, 0.57] & (0.1, 0.2] &        1 & 1 \\
Alès &           (0.19, 0.22] &           (0.66, 0.77] &  (0, 450] &  (0.37, 0.47] &       (0.2, 0.3] &        1 &        1 \\
Sin-le-Noble &           (0.19, 0.22] &           (0.66, 0.77] &  (0, 450] &  (0.47, 0.57] &       (0.2, 0.3] &        1 &        1 \\
Grand-Couronne et environs &           (0.19, 0.22] &           (0.66, 0.77] &  (0, 450] &  (0.47, 0.57] &       (0.1, 0.2] &        1 &        1 \\
Dreux &           (0.19, 0.22] &           (0.66, 0.77] &  (0, 450] &  (0.47, 0.57] &       (0.2, 0.3] &        1 &        1 \\
Vosges &           (0.19, 0.22] &           (0.77, 0.88] &  (0, 450] &  (0.47, 0.57] &       (0.1, 0.2] &        0 &        0 \\
Arras et environs &           (0.19, 0.22] &           (0.66, 0.77] &  (0, 450] &  (0.37, 0.47] &       (0.1, 0.2] &        1 &        1 \\
Montargis et environs &           (0.19, 0.22] &           (0.66, 0.77] &  (0, 450] &  (0.37, 0.47] &       (0.2, 0.3] &        1 &        1 \\
Marseille, 3e arrondissement &           (0.19, 0.22] &           (0.66, 0.77] &  (450, 900] &  (0.47, 0.57] &       (0.1, 0.2] &        1 &        1 \\
Nantes &           (0.19, 0.22] &           (0.66, 0.77] &  (0, 450] &  (0.47, 0.57] &       (0.1, 0.2] &        1 &        1 \\
Mâcon et environs &           (0.19, 0.22] &           (0.66, 0.77] &  (0, 450] &  (0.37, 0.47] &       (0.1, 0.2] &        1 &        1 \\
\bottomrule
\end{tabular}
}
\caption{Two sample matched groups generated by FLAME on the application data described in Section \ref{sec:application}. The columns are a subset of the covariates used for matching. Territory was not used for matching. Original covariates are continuous and were coarsened into 5 bins. Last election PS vote share was coarsened into 10 bins. Labels in the cells represent lower and upper bounds of the covariate bin each unit belongs to. The two groups have relatively good match quality overall.}
\label{tab:samplegroups}
\end{table*}
%\end{center}

\subsection{Sample Matched Groups}
Sample matched groups are given in Table~\ref{tab:samplegroups}. These groups were produced by \FLAMEIV on the data from \cite{pons2018}, introduced in Section \ref{sec:application}. The algorithm was ran on all of the covariates collected in the original study except for territory. Here we report some selected covariates for the groups.  The first group is comprised of electoral districts in which previous turnout was relatively good but PS vote share was low. This suggest that existing partisan splits are being taken into account by \FLAMEIV for matching. Municipalities in the second group have slightly lower turnout at the previous election but a much larger vote share for PS. Note also that treatment adoption is very high in the second group, while low in the first: this suggest that the instrument is weak in Group 1 and strong in Group 2.

\end{document}